\documentclass[longauth]{aaEC}

\usepackage{graphicx}
\usepackage{natbib}
\usepackage{scalerel}

\usepackage[table]{xcolor}

\usepackage{txfonts}
\usepackage[pdfencoding=auto,psdextra]{hyperref}
\hypersetup{
    colorlinks=true,
    linkcolor=blue,
    filecolor=magenta,      
    urlcolor=blue,
    citecolor=blue
}
\usepackage{pageslts}
\makeatletter
\renewcommand*\aa@pageof{, page \thepage{} of \pageref*{LastPage}}
\makeatother

\usepackage[utf8]{inputenc}

\usepackage{euclid}

\usepackage{csquotes}

\usepackage{amsmath}

\newcommand{\linenumbers}[0]{}

\newcommand\NeV{[\ion{Ne}{v}]$\lambda$3426} 
 
\newcommand\OII{[\ion{O}{ii}]$\lambda$3727}

\newcommand\halpha{\mbox{H\,{\sc $\alpha$}}}

\newcommand\nevno{[Ne\,\textsc{v}]}
\newcommand\OIIno{[\ion{O}{ii}]}

\defcitealias{Q1-SP003}{RW25}
\defcitealias{Q1-SP009}{SG25}
\defcitealias{Q1-SP015}{MB25}
\defcitealias{Q1-SP013}{LM25}
\defcitealias{Q1-TP005}{TM25}
\defcitealias{Q1-TP004}{RE25}
\defcitealias{Q1-SP011}{BL25}
\defcitealias{Q1-SP031}{E25}
\defcitealias{Q1-SP023}{T25} 
\defcitealias{Q1-SP027}{M25} 

\begin{document}
%
%

\title{Euclid Quick Data Release (Q1)}
\subtitle{Spectroscopic unveiling of highly ionised lines at $z=2.48$--3.88
\thanks{Dedicated to the memory of Olivier Le F{\`e}vre and Bianca Garilli, whose invaluable mentorship and guidance were instrumental to this work.}}
\newcommand{\orcid}[1]{} 
\author{Euclid Collaboration: D.~Vergani\orcid{0000-0003-0898-2216}\thanks{\email{daniela.vergani@inaf.it}}\inst{\ref{aff1}}
\and S.~Quai\orcid{0000-0002-0449-8163}\inst{\ref{aff2},\ref{aff1}}
\and F.~Ricci\orcid{0000-0001-5742-5980}\inst{\ref{aff3},\ref{aff4}}
\and Y.~Fu\orcid{0000-0002-0759-0504}\inst{\ref{aff5},\ref{aff6}}
\and S.~Serjeant\orcid{0000-0002-0517-7943}\inst{\ref{aff7}}
\and M.~Salvato\orcid{0000-0001-7116-9303}\inst{\ref{aff8}}
\and W.~Roster\orcid{0000-0002-9149-6528}\inst{\ref{aff8}}
\and M.~Mezcua\orcid{0000-0003-4440-259X}\inst{\ref{aff9},\ref{aff10}}
\and M.~Siudek\orcid{0000-0002-2949-2155}\inst{\ref{aff11},\ref{aff9}}
\and A.~Enia\orcid{0000-0002-0200-2857}\inst{\ref{aff1},\ref{aff12}}
\and G.~Zamorani\orcid{0000-0002-2318-301X}\inst{\ref{aff1}}
\and L.~Bisigello\orcid{0000-0003-0492-4924}\inst{\ref{aff13}}
\and A.~Feltre\orcid{0000-0001-6865-2871}\inst{\ref{aff14}}
\and S.~Fotopoulou\orcid{0000-0002-9686-254X}\inst{\ref{aff15}}
\and T.~Matamoro~Zatarain\orcid{0009-0007-2976-293X}\inst{\ref{aff15}}
\and L.~Pozzetti\orcid{0000-0001-7085-0412}\inst{\ref{aff1}}
\and D.~Scott\orcid{0000-0002-6878-9840}\inst{\ref{aff16}}
\and B.~Laloux\orcid{0000-0001-9996-9732}\inst{\ref{aff17},\ref{aff8}}
\and J.~G.~Sorce\orcid{0000-0002-2307-2432}\inst{\ref{aff18},\ref{aff19}}
\and P.~A.~C.~Cunha\orcid{0000-0002-9454-859X}\inst{\ref{aff12},\ref{aff1}}
\and A.~Viitanen\orcid{0000-0001-9383-786X}\inst{\ref{aff20},\ref{aff21},\ref{aff4}}
\and C.~Saulder\orcid{0000-0002-0408-5633}\inst{\ref{aff8},\ref{aff22}}
\and E.~Rossetti\orcid{0000-0003-0238-4047}\inst{\ref{aff12}}
\and M.~Moresco\orcid{0000-0002-7616-7136}\inst{\ref{aff2},\ref{aff1}}
\and V.~Le~Brun\orcid{0000-0002-5027-1939}\inst{\ref{aff23}}
\and E.~Palazzi\orcid{0000-0002-8691-7666}\inst{\ref{aff1}}
\and M.~Talia\orcid{0000-0003-4352-2063}\inst{\ref{aff2},\ref{aff1}}
\and Z.~Mao\orcid{0000-0003-4016-845X}\inst{\ref{aff1}}
\and L.~Nicastro\orcid{0000-0001-8534-6788}\inst{\ref{aff1}}
\and E.~Maiorano\orcid{0000-0003-2593-4355}\inst{\ref{aff1}}
\and D.~Vibert\orcid{0009-0008-0607-631X}\inst{\ref{aff23}}
\and P.-Y.~Chabaud\inst{\ref{aff23}}
\and G.~Daste\inst{\ref{aff23}}
\and F.~Dufresne\inst{\ref{aff23}}
\and T.~Bedrine\inst{\ref{aff23}}
\and A.~Amara\inst{\ref{aff24}}
\and S.~Andreon\orcid{0000-0002-2041-8784}\inst{\ref{aff25}}
\and N.~Auricchio\orcid{0000-0003-4444-8651}\inst{\ref{aff1}}
\and C.~Baccigalupi\orcid{0000-0002-8211-1630}\inst{\ref{aff26},\ref{aff27},\ref{aff28},\ref{aff29}}
\and M.~Baldi\orcid{0000-0003-4145-1943}\inst{\ref{aff12},\ref{aff1},\ref{aff30}}
\and A.~Balestra\orcid{0000-0002-6967-261X}\inst{\ref{aff13}}
\and S.~Bardelli\orcid{0000-0002-8900-0298}\inst{\ref{aff1}}
\and P.~Battaglia\orcid{0000-0002-7337-5909}\inst{\ref{aff1}}
\and A.~Biviano\orcid{0000-0002-0857-0732}\inst{\ref{aff27},\ref{aff26}}
\and E.~Branchini\orcid{0000-0002-0808-6908}\inst{\ref{aff31},\ref{aff32},\ref{aff25}}
\and M.~Brescia\orcid{0000-0001-9506-5680}\inst{\ref{aff33},\ref{aff17}}
\and S.~Camera\orcid{0000-0003-3399-3574}\inst{\ref{aff34},\ref{aff35},\ref{aff36}}
\and G.~Ca\~nas-Herrera\orcid{0000-0003-2796-2149}\inst{\ref{aff37},\ref{aff5}}
\and V.~Capobianco\orcid{0000-0002-3309-7692}\inst{\ref{aff36}}
\and C.~Carbone\orcid{0000-0003-0125-3563}\inst{\ref{aff38}}
\and J.~Carretero\orcid{0000-0002-3130-0204}\inst{\ref{aff39},\ref{aff40}}
\and S.~Casas\orcid{0000-0002-4751-5138}\inst{\ref{aff41},\ref{aff42}}
\and M.~Castellano\orcid{0000-0001-9875-8263}\inst{\ref{aff4}}
\and G.~Castignani\orcid{0000-0001-6831-0687}\inst{\ref{aff1}}
\and S.~Cavuoti\orcid{0000-0002-3787-4196}\inst{\ref{aff17},\ref{aff43}}
\and K.~C.~Chambers\orcid{0000-0001-6965-7789}\inst{\ref{aff44}}
\and A.~Cimatti\inst{\ref{aff45}}
\and C.~Colodro-Conde\inst{\ref{aff46}}
\and G.~Congedo\orcid{0000-0003-2508-0046}\inst{\ref{aff47}}
\and C.~J.~Conselice\orcid{0000-0003-1949-7638}\inst{\ref{aff48}}
\and L.~Conversi\orcid{0000-0002-6710-8476}\inst{\ref{aff49},\ref{aff50}}
\and Y.~Copin\orcid{0000-0002-5317-7518}\inst{\ref{aff51}}
\and F.~Courbin\orcid{0000-0003-0758-6510}\inst{\ref{aff52},\ref{aff53},\ref{aff54}}
\and H.~M.~Courtois\orcid{0000-0003-0509-1776}\inst{\ref{aff55}}
\and A.~Da~Silva\orcid{0000-0002-6385-1609}\inst{\ref{aff56},\ref{aff57}}
\and H.~Degaudenzi\orcid{0000-0002-5887-6799}\inst{\ref{aff21}}
\and G.~De~Lucia\orcid{0000-0002-6220-9104}\inst{\ref{aff27}}
\and A.~M.~Di~Giorgio\orcid{0000-0002-4767-2360}\inst{\ref{aff58}}
\and H.~Dole\orcid{0000-0002-9767-3839}\inst{\ref{aff19}}
\and F.~Dubath\orcid{0000-0002-6533-2810}\inst{\ref{aff21}}
\and X.~Dupac\inst{\ref{aff50}}
\and S.~Dusini\orcid{0000-0002-1128-0664}\inst{\ref{aff59}}
\and S.~Escoffier\orcid{0000-0002-2847-7498}\inst{\ref{aff60}}
\and M.~Farina\orcid{0000-0002-3089-7846}\inst{\ref{aff58}}
\and R.~Farinelli\inst{\ref{aff1}}
\and F.~Faustini\orcid{0000-0001-6274-5145}\inst{\ref{aff4},\ref{aff61}}
\and S.~Ferriol\inst{\ref{aff51}}
\and F.~Finelli\orcid{0000-0002-6694-3269}\inst{\ref{aff1},\ref{aff62}}
\and N.~Fourmanoit\orcid{0009-0005-6816-6925}\inst{\ref{aff60}}
\and M.~Frailis\orcid{0000-0002-7400-2135}\inst{\ref{aff27}}
\and E.~Franceschi\orcid{0000-0002-0585-6591}\inst{\ref{aff1}}
\and M.~Fumana\orcid{0000-0001-6787-5950}\inst{\ref{aff38}}
\and S.~Galeotta\orcid{0000-0002-3748-5115}\inst{\ref{aff27}}
\and W.~Gillard\orcid{0000-0003-4744-9748}\inst{\ref{aff60}}
\and B.~Gillis\orcid{0000-0002-4478-1270}\inst{\ref{aff47}}
\and C.~Giocoli\orcid{0000-0002-9590-7961}\inst{\ref{aff1},\ref{aff30}}
\and J.~Gracia-Carpio\inst{\ref{aff8}}
\and A.~Grazian\orcid{0000-0002-5688-0663}\inst{\ref{aff13}}
\and F.~Grupp\inst{\ref{aff8},\ref{aff22}}
\and S.~V.~H.~Haugan\orcid{0000-0001-9648-7260}\inst{\ref{aff63}}
\and J.~Hoar\inst{\ref{aff50}}
\and H.~Hoekstra\orcid{0000-0002-0641-3231}\inst{\ref{aff5}}
\and W.~Holmes\inst{\ref{aff64}}
\and I.~M.~Hook\orcid{0000-0002-2960-978X}\inst{\ref{aff65}}
\and F.~Hormuth\inst{\ref{aff66}}
\and A.~Hornstrup\orcid{0000-0002-3363-0936}\inst{\ref{aff67},\ref{aff68}}
\and K.~Jahnke\orcid{0000-0003-3804-2137}\inst{\ref{aff69}}
\and M.~Jhabvala\inst{\ref{aff70}}
\and E.~Keih\"anen\orcid{0000-0003-1804-7715}\inst{\ref{aff20}}
\and S.~Kermiche\orcid{0000-0002-0302-5735}\inst{\ref{aff60}}
\and A.~Kiessling\orcid{0000-0002-2590-1273}\inst{\ref{aff64}}
\and B.~Kubik\orcid{0009-0006-5823-4880}\inst{\ref{aff51}}
\and M.~K\"ummel\orcid{0000-0003-2791-2117}\inst{\ref{aff22}}
\and M.~Kunz\orcid{0000-0002-3052-7394}\inst{\ref{aff71}}
\and H.~Kurki-Suonio\orcid{0000-0002-4618-3063}\inst{\ref{aff72},\ref{aff73}}
\and A.~M.~C.~Le~Brun\orcid{0000-0002-0936-4594}\inst{\ref{aff74}}
\and D.~Le~Mignant\orcid{0000-0002-5339-5515}\inst{\ref{aff23}}
\and S.~Ligori\orcid{0000-0003-4172-4606}\inst{\ref{aff36}}
\and P.~B.~Lilje\orcid{0000-0003-4324-7794}\inst{\ref{aff63}}
\and V.~Lindholm\orcid{0000-0003-2317-5471}\inst{\ref{aff72},\ref{aff73}}
\and I.~Lloro\orcid{0000-0001-5966-1434}\inst{\ref{aff75}}
\and G.~Mainetti\orcid{0000-0003-2384-2377}\inst{\ref{aff76}}
\and D.~Maino\inst{\ref{aff77},\ref{aff38},\ref{aff78}}
\and O.~Mansutti\orcid{0000-0001-5758-4658}\inst{\ref{aff27}}
\and O.~Marggraf\orcid{0000-0001-7242-3852}\inst{\ref{aff79}}
\and M.~Martinelli\orcid{0000-0002-6943-7732}\inst{\ref{aff4},\ref{aff80}}
\and N.~Martinet\orcid{0000-0003-2786-7790}\inst{\ref{aff23}}
\and F.~Marulli\orcid{0000-0002-8850-0303}\inst{\ref{aff2},\ref{aff1},\ref{aff30}}
\and R.~J.~Massey\orcid{0000-0002-6085-3780}\inst{\ref{aff81}}
\and E.~Medinaceli\orcid{0000-0002-4040-7783}\inst{\ref{aff1}}
\and S.~Mei\orcid{0000-0002-2849-559X}\inst{\ref{aff82},\ref{aff83}}
\and M.~Melchior\inst{\ref{aff84}}
\and Y.~Mellier\inst{\ref{aff85},\ref{aff86}}
\and M.~Meneghetti\orcid{0000-0003-1225-7084}\inst{\ref{aff1},\ref{aff30}}
\and E.~Merlin\orcid{0000-0001-6870-8900}\inst{\ref{aff4}}
\and G.~Meylan\inst{\ref{aff87}}
\and A.~Mora\orcid{0000-0002-1922-8529}\inst{\ref{aff88}}
\and L.~Moscardini\orcid{0000-0002-3473-6716}\inst{\ref{aff2},\ref{aff1},\ref{aff30}}
\and C.~Neissner\orcid{0000-0001-8524-4968}\inst{\ref{aff89},\ref{aff40}}
\and S.-M.~Niemi\orcid{0009-0005-0247-0086}\inst{\ref{aff37}}
\and C.~Padilla\orcid{0000-0001-7951-0166}\inst{\ref{aff89}}
\and S.~Paltani\orcid{0000-0002-8108-9179}\inst{\ref{aff21}}
\and F.~Pasian\orcid{0000-0002-4869-3227}\inst{\ref{aff27}}
\and K.~Pedersen\inst{\ref{aff90}}
\and W.~J.~Percival\orcid{0000-0002-0644-5727}\inst{\ref{aff91},\ref{aff92},\ref{aff93}}
\and V.~Pettorino\orcid{0000-0002-4203-9320}\inst{\ref{aff37}}
\and S.~Pires\orcid{0000-0002-0249-2104}\inst{\ref{aff94}}
\and G.~Polenta\orcid{0000-0003-4067-9196}\inst{\ref{aff61}}
\and M.~Poncet\inst{\ref{aff95}}
\and L.~A.~Popa\inst{\ref{aff96}}
\and F.~Raison\orcid{0000-0002-7819-6918}\inst{\ref{aff8}}
\and R.~Rebolo\orcid{0000-0003-3767-7085}\inst{\ref{aff46},\ref{aff97},\ref{aff98}}
\and A.~Renzi\orcid{0000-0001-9856-1970}\inst{\ref{aff99},\ref{aff59}}
\and J.~Rhodes\orcid{0000-0002-4485-8549}\inst{\ref{aff64}}
\and G.~Riccio\inst{\ref{aff17}}
\and E.~Romelli\orcid{0000-0003-3069-9222}\inst{\ref{aff27}}
\and M.~Roncarelli\orcid{0000-0001-9587-7822}\inst{\ref{aff1}}
\and R.~Saglia\orcid{0000-0003-0378-7032}\inst{\ref{aff22},\ref{aff8}}
\and Z.~Sakr\orcid{0000-0002-4823-3757}\inst{\ref{aff100},\ref{aff101},\ref{aff102}}
\and D.~Sapone\orcid{0000-0001-7089-4503}\inst{\ref{aff103}}
\and B.~Sartoris\orcid{0000-0003-1337-5269}\inst{\ref{aff22},\ref{aff27}}
\and M.~Schirmer\orcid{0000-0003-2568-9994}\inst{\ref{aff69}}
\and P.~Schneider\orcid{0000-0001-8561-2679}\inst{\ref{aff79}}
\and T.~Schrabback\orcid{0000-0002-6987-7834}\inst{\ref{aff104}}
\and M.~Scodeggio\inst{\ref{aff38}}
\and A.~Secroun\orcid{0000-0003-0505-3710}\inst{\ref{aff60}}
\and G.~Seidel\orcid{0000-0003-2907-353X}\inst{\ref{aff69}}
\and S.~Serrano\orcid{0000-0002-0211-2861}\inst{\ref{aff10},\ref{aff105},\ref{aff9}}
\and P.~Simon\inst{\ref{aff79}}
\and C.~Sirignano\orcid{0000-0002-0995-7146}\inst{\ref{aff99},\ref{aff59}}
\and G.~Sirri\orcid{0000-0003-2626-2853}\inst{\ref{aff30}}
\and L.~Stanco\orcid{0000-0002-9706-5104}\inst{\ref{aff59}}
\and J.~Steinwagner\orcid{0000-0001-7443-1047}\inst{\ref{aff8}}
\and C.~Surace\orcid{0000-0003-2592-0113}\inst{\ref{aff23}}
\and P.~Tallada-Cresp\'{i}\orcid{0000-0002-1336-8328}\inst{\ref{aff39},\ref{aff40}}
\and A.~N.~Taylor\inst{\ref{aff47}}
\and H.~I.~Teplitz\orcid{0000-0002-7064-5424}\inst{\ref{aff106}}
\and I.~Tereno\orcid{0000-0002-4537-6218}\inst{\ref{aff56},\ref{aff107}}
\and N.~Tessore\orcid{0000-0002-9696-7931}\inst{\ref{aff108},\ref{aff109}}
\and S.~Toft\orcid{0000-0003-3631-7176}\inst{\ref{aff110},\ref{aff111}}
\and R.~Toledo-Moreo\orcid{0000-0002-2997-4859}\inst{\ref{aff112}}
\and F.~Torradeflot\orcid{0000-0003-1160-1517}\inst{\ref{aff40},\ref{aff39}}
\and I.~Tutusaus\orcid{0000-0002-3199-0399}\inst{\ref{aff9},\ref{aff10},\ref{aff101}}
\and L.~Valenziano\orcid{0000-0002-1170-0104}\inst{\ref{aff1},\ref{aff62}}
\and J.~Valiviita\orcid{0000-0001-6225-3693}\inst{\ref{aff72},\ref{aff73}}
\and T.~Vassallo\orcid{0000-0001-6512-6358}\inst{\ref{aff27}}
\and A.~Veropalumbo\orcid{0000-0003-2387-1194}\inst{\ref{aff25},\ref{aff32},\ref{aff31}}
\and Y.~Wang\orcid{0000-0002-4749-2984}\inst{\ref{aff106}}
\and J.~Weller\orcid{0000-0002-8282-2010}\inst{\ref{aff22},\ref{aff8}}
\and A.~Zacchei\orcid{0000-0003-0396-1192}\inst{\ref{aff27},\ref{aff26}}
\and F.~M.~Zerbi\inst{\ref{aff25}}
\and E.~Zucca\orcid{0000-0002-5845-8132}\inst{\ref{aff1}}
\and V.~Allevato\orcid{0000-0001-7232-5152}\inst{\ref{aff17}}
\and M.~Ballardini\orcid{0000-0003-4481-3559}\inst{\ref{aff113},\ref{aff114},\ref{aff1}}
\and M.~Bolzonella\orcid{0000-0003-3278-4607}\inst{\ref{aff1}}
\and E.~Bozzo\orcid{0000-0002-8201-1525}\inst{\ref{aff21}}
\and C.~Burigana\orcid{0000-0002-3005-5796}\inst{\ref{aff115},\ref{aff62}}
\and R.~Cabanac\orcid{0000-0001-6679-2600}\inst{\ref{aff101}}
\and M.~Calabrese\orcid{0000-0002-2637-2422}\inst{\ref{aff116},\ref{aff38}}
\and A.~Cappi\inst{\ref{aff1},\ref{aff117}}
\and D.~Di~Ferdinando\inst{\ref{aff30}}
\and J.~A.~Escartin~Vigo\inst{\ref{aff8}}
\and L.~Gabarra\orcid{0000-0002-8486-8856}\inst{\ref{aff118}}
\and W.~G.~Hartley\inst{\ref{aff21}}
\and M.~Huertas-Company\orcid{0000-0002-1416-8483}\inst{\ref{aff46},\ref{aff11},\ref{aff119},\ref{aff120}}
\and R.~Maoli\orcid{0000-0002-6065-3025}\inst{\ref{aff121},\ref{aff4}}
\and J.~Mart\'{i}n-Fleitas\orcid{0000-0002-8594-569X}\inst{\ref{aff122}}
\and S.~Matthew\orcid{0000-0001-8448-1697}\inst{\ref{aff47}}
\and N.~Mauri\orcid{0000-0001-8196-1548}\inst{\ref{aff45},\ref{aff30}}
\and R.~B.~Metcalf\orcid{0000-0003-3167-2574}\inst{\ref{aff2},\ref{aff1}}
\and A.~Pezzotta\orcid{0000-0003-0726-2268}\inst{\ref{aff25}}
\and M.~P\"ontinen\orcid{0000-0001-5442-2530}\inst{\ref{aff72}}
\and C.~Porciani\orcid{0000-0002-7797-2508}\inst{\ref{aff79}}
\and I.~Risso\orcid{0000-0003-2525-7761}\inst{\ref{aff25},\ref{aff32}}
\and V.~Scottez\orcid{0009-0008-3864-940X}\inst{\ref{aff85},\ref{aff123}}
\and M.~Sereno\orcid{0000-0003-0302-0325}\inst{\ref{aff1},\ref{aff30}}
\and M.~Tenti\orcid{0000-0002-4254-5901}\inst{\ref{aff30}}
\and M.~Viel\orcid{0000-0002-2642-5707}\inst{\ref{aff26},\ref{aff27},\ref{aff29},\ref{aff28},\ref{aff124}}
\and M.~Wiesmann\orcid{0009-0000-8199-5860}\inst{\ref{aff63}}
\and Y.~Akrami\orcid{0000-0002-2407-7956}\inst{\ref{aff125},\ref{aff126}}
\and I.~T.~Andika\orcid{0000-0001-6102-9526}\inst{\ref{aff127},\ref{aff128}}
\and S.~Anselmi\orcid{0000-0002-3579-9583}\inst{\ref{aff59},\ref{aff99},\ref{aff129}}
\and M.~Archidiacono\orcid{0000-0003-4952-9012}\inst{\ref{aff77},\ref{aff78}}
\and F.~Atrio-Barandela\orcid{0000-0002-2130-2513}\inst{\ref{aff130}}
\and D.~Bertacca\orcid{0000-0002-2490-7139}\inst{\ref{aff99},\ref{aff13},\ref{aff59}}
\and M.~Bethermin\orcid{0000-0002-3915-2015}\inst{\ref{aff131}}
\and A.~Blanchard\orcid{0000-0001-8555-9003}\inst{\ref{aff101}}
\and L.~Blot\orcid{0000-0002-9622-7167}\inst{\ref{aff132},\ref{aff74}}
\and M.~Bonici\orcid{0000-0002-8430-126X}\inst{\ref{aff91},\ref{aff38}}
\and S.~Borgani\orcid{0000-0001-6151-6439}\inst{\ref{aff133},\ref{aff26},\ref{aff27},\ref{aff28},\ref{aff124}}
\and M.~L.~Brown\orcid{0000-0002-0370-8077}\inst{\ref{aff48}}
\and S.~Bruton\orcid{0000-0002-6503-5218}\inst{\ref{aff134}}
\and A.~Calabro\orcid{0000-0003-2536-1614}\inst{\ref{aff4}}
\and B.~Camacho~Quevedo\orcid{0000-0002-8789-4232}\inst{\ref{aff26},\ref{aff29},\ref{aff27}}
\and F.~Caro\inst{\ref{aff4}}
\and C.~S.~Carvalho\inst{\ref{aff107}}
\and T.~Castro\orcid{0000-0002-6292-3228}\inst{\ref{aff27},\ref{aff28},\ref{aff26},\ref{aff124}}
\and F.~Cogato\orcid{0000-0003-4632-6113}\inst{\ref{aff2},\ref{aff1}}
\and S.~Conseil\orcid{0000-0002-3657-4191}\inst{\ref{aff51}}
\and T.~Contini\orcid{0000-0003-0275-938X}\inst{\ref{aff101}}
\and A.~R.~Cooray\orcid{0000-0002-3892-0190}\inst{\ref{aff135}}
\and O.~Cucciati\orcid{0000-0002-9336-7551}\inst{\ref{aff1}}
\and G.~Desprez\orcid{0000-0001-8325-1742}\inst{\ref{aff6}}
\and A.~D\'iaz-S\'anchez\orcid{0000-0003-0748-4768}\inst{\ref{aff136}}
\and J.~J.~Diaz\orcid{0000-0003-2101-1078}\inst{\ref{aff46}}
\and S.~Di~Domizio\orcid{0000-0003-2863-5895}\inst{\ref{aff31},\ref{aff32}}
\and J.~M.~Diego\orcid{0000-0001-9065-3926}\inst{\ref{aff137}}
\and M.~Y.~Elkhashab\orcid{0000-0001-9306-2603}\inst{\ref{aff27},\ref{aff28},\ref{aff133},\ref{aff26}}
\and Y.~Fang\orcid{0000-0002-0334-6950}\inst{\ref{aff22}}
\and A.~Finoguenov\orcid{0000-0002-4606-5403}\inst{\ref{aff72}}
\and F.~Fontanot\orcid{0000-0003-4744-0188}\inst{\ref{aff27},\ref{aff26}}
\and A.~Franco\orcid{0000-0002-4761-366X}\inst{\ref{aff138},\ref{aff139},\ref{aff140}}
\and K.~Ganga\orcid{0000-0001-8159-8208}\inst{\ref{aff82}}
\and J.~Garc\'ia-Bellido\orcid{0000-0002-9370-8360}\inst{\ref{aff125}}
\and T.~Gasparetto\orcid{0000-0002-7913-4866}\inst{\ref{aff4}}
\and V.~Gautard\inst{\ref{aff141}}
\and E.~Gaztanaga\orcid{0000-0001-9632-0815}\inst{\ref{aff9},\ref{aff10},\ref{aff142}}
\and F.~Giacomini\orcid{0000-0002-3129-2814}\inst{\ref{aff30}}
\and F.~Gianotti\orcid{0000-0003-4666-119X}\inst{\ref{aff1}}
\and G.~Gozaliasl\orcid{0000-0002-0236-919X}\inst{\ref{aff143},\ref{aff72}}
\and M.~Guidi\orcid{0000-0001-9408-1101}\inst{\ref{aff12},\ref{aff1}}
\and C.~M.~Gutierrez\orcid{0000-0001-7854-783X}\inst{\ref{aff11}}
\and A.~Hall\orcid{0000-0002-3139-8651}\inst{\ref{aff47}}
\and S.~Hemmati\orcid{0000-0003-2226-5395}\inst{\ref{aff144}}
\and C.~Hern\'andez-Monteagudo\orcid{0000-0001-5471-9166}\inst{\ref{aff98},\ref{aff46}}
\and H.~Hildebrandt\orcid{0000-0002-9814-3338}\inst{\ref{aff145}}
\and J.~Hjorth\orcid{0000-0002-4571-2306}\inst{\ref{aff90}}
\and J.~J.~E.~Kajava\orcid{0000-0002-3010-8333}\inst{\ref{aff146},\ref{aff147}}
\and Y.~Kang\orcid{0009-0000-8588-7250}\inst{\ref{aff21}}
\and V.~Kansal\orcid{0000-0002-4008-6078}\inst{\ref{aff148},\ref{aff149}}
\and D.~Karagiannis\orcid{0000-0002-4927-0816}\inst{\ref{aff113},\ref{aff150}}
\and K.~Kiiveri\inst{\ref{aff20}}
\and J.~Kim\orcid{0000-0003-2776-2761}\inst{\ref{aff118}}
\and C.~C.~Kirkpatrick\inst{\ref{aff20}}
\and S.~Kruk\orcid{0000-0001-8010-8879}\inst{\ref{aff50}}
\and M.~Lattanzi\orcid{0000-0003-1059-2532}\inst{\ref{aff114}}
\and L.~Legrand\orcid{0000-0003-0610-5252}\inst{\ref{aff151},\ref{aff152}}
\and M.~Lembo\orcid{0000-0002-5271-5070}\inst{\ref{aff86},\ref{aff113},\ref{aff114}}
\and F.~Lepori\orcid{0009-0000-5061-7138}\inst{\ref{aff153}}
\and G.~Leroy\orcid{0009-0004-2523-4425}\inst{\ref{aff154},\ref{aff81}}
\and G.~F.~Lesci\orcid{0000-0002-4607-2830}\inst{\ref{aff2},\ref{aff1}}
\and J.~Lesgourgues\orcid{0000-0001-7627-353X}\inst{\ref{aff41}}
\and T.~I.~Liaudat\orcid{0000-0002-9104-314X}\inst{\ref{aff155}}
\and A.~Loureiro\orcid{0000-0002-4371-0876}\inst{\ref{aff156},\ref{aff157}}
\and J.~Macias-Perez\orcid{0000-0002-5385-2763}\inst{\ref{aff158}}
\and M.~Magliocchetti\orcid{0000-0001-9158-4838}\inst{\ref{aff58}}
\and C.~Mancini\orcid{0000-0002-4297-0561}\inst{\ref{aff38}}
\and F.~Mannucci\orcid{0000-0002-4803-2381}\inst{\ref{aff14}}
\and C.~J.~A.~P.~Martins\orcid{0000-0002-4886-9261}\inst{\ref{aff159},\ref{aff160}}
\and L.~Maurin\orcid{0000-0002-8406-0857}\inst{\ref{aff19}}
\and M.~Miluzio\inst{\ref{aff50},\ref{aff161}}
\and P.~Monaco\orcid{0000-0003-2083-7564}\inst{\ref{aff133},\ref{aff27},\ref{aff28},\ref{aff26}}
\and C.~Moretti\orcid{0000-0003-3314-8936}\inst{\ref{aff27},\ref{aff26},\ref{aff28},\ref{aff29}}
\and G.~Morgante\inst{\ref{aff1}}
\and S.~Nadathur\orcid{0000-0001-9070-3102}\inst{\ref{aff142}}
\and K.~Naidoo\orcid{0000-0002-9182-1802}\inst{\ref{aff142},\ref{aff108}}
\and A.~Navarro-Alsina\orcid{0000-0002-3173-2592}\inst{\ref{aff79}}
\and S.~Nesseris\orcid{0000-0002-0567-0324}\inst{\ref{aff125}}
\and D.~Paoletti\orcid{0000-0003-4761-6147}\inst{\ref{aff1},\ref{aff62}}
\and F.~Passalacqua\orcid{0000-0002-8606-4093}\inst{\ref{aff99},\ref{aff59}}
\and K.~Paterson\orcid{0000-0001-8340-3486}\inst{\ref{aff69}}
\and L.~Patrizii\inst{\ref{aff30}}
\and A.~Pisani\orcid{0000-0002-6146-4437}\inst{\ref{aff60}}
\and D.~Potter\orcid{0000-0002-0757-5195}\inst{\ref{aff153}}
\and M.~Radovich\orcid{0000-0002-3585-866X}\inst{\ref{aff13}}
\and G.~Rodighiero\orcid{0000-0002-9415-2296}\inst{\ref{aff99},\ref{aff13}}
\and S.~Sacquegna\orcid{0000-0002-8433-6630}\inst{\ref{aff162}}
\and M.~Sahl\'en\orcid{0000-0003-0973-4804}\inst{\ref{aff163}}
\and D.~B.~Sanders\orcid{0000-0002-1233-9998}\inst{\ref{aff44}}
\and E.~Sarpa\orcid{0000-0002-1256-655X}\inst{\ref{aff29},\ref{aff124},\ref{aff28}}
\and C.~Scarlata\orcid{0000-0002-9136-8876}\inst{\ref{aff164}}
\and A.~Schneider\orcid{0000-0001-7055-8104}\inst{\ref{aff153}}
\and D.~Sciotti\orcid{0009-0008-4519-2620}\inst{\ref{aff4},\ref{aff80}}
\and E.~Sellentin\inst{\ref{aff165},\ref{aff5}}
\and F.~Shankar\orcid{0000-0001-8973-5051}\inst{\ref{aff166}}
\and L.~C.~Smith\orcid{0000-0002-3259-2771}\inst{\ref{aff167}}
\and K.~Tanidis\orcid{0000-0001-9843-5130}\inst{\ref{aff118}}
\and C.~Tao\orcid{0000-0001-7961-8177}\inst{\ref{aff60}}
\and G.~Testera\inst{\ref{aff32}}
\and R.~Teyssier\orcid{0000-0001-7689-0933}\inst{\ref{aff168}}
\and S.~Tosi\orcid{0000-0002-7275-9193}\inst{\ref{aff31},\ref{aff32},\ref{aff25}}
\and A.~Troja\orcid{0000-0003-0239-4595}\inst{\ref{aff99},\ref{aff59}}
\and M.~Tucci\inst{\ref{aff21}}
\and C.~Valieri\inst{\ref{aff30}}
\and A.~Venhola\orcid{0000-0001-6071-4564}\inst{\ref{aff169}}
\and G.~Verza\orcid{0000-0002-1886-8348}\inst{\ref{aff170}}
\and P.~Vielzeuf\orcid{0000-0003-2035-9339}\inst{\ref{aff60}}
\and N.~A.~Walton\orcid{0000-0003-3983-8778}\inst{\ref{aff167}}}
										   
\institute{INAF-Osservatorio di Astrofisica e Scienza dello Spazio di Bologna, Via Piero Gobetti 93/3, 40129 Bologna, Italy\label{aff1}
\and
Dipartimento di Fisica e Astronomia "Augusto Righi" - Alma Mater Studiorum Universit\`a di Bologna, via Piero Gobetti 93/2, 40129 Bologna, Italy\label{aff2}
\and
Department of Mathematics and Physics, Roma Tre University, Via della Vasca Navale 84, 00146 Rome, Italy\label{aff3}
\and
INAF-Osservatorio Astronomico di Roma, Via Frascati 33, 00078 Monteporzio Catone, Italy\label{aff4}
\and
Leiden Observatory, Leiden University, Einsteinweg 55, 2333 CC Leiden, The Netherlands\label{aff5}
\and
Kapteyn Astronomical Institute, University of Groningen, PO Box 800, 9700 AV Groningen, The Netherlands\label{aff6}
\and
School of Physical Sciences, The Open University, Milton Keynes, MK7 6AA, UK\label{aff7}
\and
Max Planck Institute for Extraterrestrial Physics, Giessenbachstr. 1, 85748 Garching, Germany\label{aff8}
\and
Institute of Space Sciences (ICE, CSIC), Campus UAB, Carrer de Can Magrans, s/n, 08193 Barcelona, Spain\label{aff9}
\and
Institut d'Estudis Espacials de Catalunya (IEEC),  Edifici RDIT, Campus UPC, 08860 Castelldefels, Barcelona, Spain\label{aff10}
\and
 Instituto de Astrof\'{\i}sica de Canarias, E-38205 La Laguna; Universidad de La Laguna, Dpto. Astrof\'\i sica, E-38206 La Laguna, Tenerife, Spain\label{aff11}
\and
Dipartimento di Fisica e Astronomia, Universit\`a di Bologna, Via Gobetti 93/2, 40129 Bologna, Italy\label{aff12}
\and
INAF-Osservatorio Astronomico di Padova, Via dell'Osservatorio 5, 35122 Padova, Italy\label{aff13}
\and
INAF-Osservatorio Astrofisico di Arcetri, Largo E. Fermi 5, 50125, Firenze, Italy\label{aff14}
\and
School of Physics, HH Wills Physics Laboratory, University of Bristol, Tyndall Avenue, Bristol, BS8 1TL, UK\label{aff15}
\and
Department of Physics and Astronomy, University of British Columbia, Vancouver, BC V6T 1Z1, Canada\label{aff16}
\and
INAF-Osservatorio Astronomico di Capodimonte, Via Moiariello 16, 80131 Napoli, Italy\label{aff17}
\and
Univ. Lille, CNRS, Centrale Lille, UMR 9189 CRIStAL, 59000 Lille, France\label{aff18}
\and
Universit\'e Paris-Saclay, CNRS, Institut d'astrophysique spatiale, 91405, Orsay, France\label{aff19}
\and
Department of Physics and Helsinki Institute of Physics, Gustaf H\"allstr\"omin katu 2, University of Helsinki, 00014 Helsinki, Finland\label{aff20}
\and
Department of Astronomy, University of Geneva, ch. d'Ecogia 16, 1290 Versoix, Switzerland\label{aff21}
\and
Universit\"ats-Sternwarte M\"unchen, Fakult\"at f\"ur Physik, Ludwig-Maximilians-Universit\"at M\"unchen, Scheinerstr.~1, 81679 M\"unchen, Germany\label{aff22}
\and
Aix-Marseille Universit\'e, CNRS, CNES, LAM, Marseille, France\label{aff23}
\and
School of Mathematics and Physics, University of Surrey, Guildford, Surrey, GU2 7XH, UK\label{aff24}
\and
INAF-Osservatorio Astronomico di Brera, Via Brera 28, 20122 Milano, Italy\label{aff25}
\and
IFPU, Institute for Fundamental Physics of the Universe, via Beirut 2, 34151 Trieste, Italy\label{aff26}
\and
INAF-Osservatorio Astronomico di Trieste, Via G. B. Tiepolo 11, 34143 Trieste, Italy\label{aff27}
\and
INFN, Sezione di Trieste, Via Valerio 2, 34127 Trieste TS, Italy\label{aff28}
\and
SISSA, International School for Advanced Studies, Via Bonomea 265, 34136 Trieste TS, Italy\label{aff29}
\and
INFN-Sezione di Bologna, Viale Berti Pichat 6/2, 40127 Bologna, Italy\label{aff30}
\and
Dipartimento di Fisica, Universit\`a di Genova, Via Dodecaneso 33, 16146, Genova, Italy\label{aff31}
\and
INFN-Sezione di Genova, Via Dodecaneso 33, 16146, Genova, Italy\label{aff32}
\and
Department of Physics "E. Pancini", University Federico II, Via Cinthia 6, 80126, Napoli, Italy\label{aff33}
\and
Dipartimento di Fisica, Universit\`a degli Studi di Torino, Via P. Giuria 1, 10125 Torino, Italy\label{aff34}
\and
INFN-Sezione di Torino, Via P. Giuria 1, 10125 Torino, Italy\label{aff35}
\and
INAF-Osservatorio Astrofisico di Torino, Via Osservatorio 20, 10025 Pino Torinese (TO), Italy\label{aff36}
\and
European Space Agency/ESTEC, Keplerlaan 1, 2201 AZ Noordwijk, The Netherlands\label{aff37}
\and
INAF-IASF Milano, Via Alfonso Corti 12, 20133 Milano, Italy\label{aff38}
\and
Centro de Investigaciones Energ\'eticas, Medioambientales y Tecnol\'ogicas (CIEMAT), Avenida Complutense 40, 28040 Madrid, Spain\label{aff39}
\and
Port d'Informaci\'{o} Cient\'{i}fica, Campus UAB, C. Albareda s/n, 08193 Bellaterra (Barcelona), Spain\label{aff40}
\and
Institute for Theoretical Particle Physics and Cosmology (TTK), RWTH Aachen University, 52056 Aachen, Germany\label{aff41}
\and
Deutsches Zentrum f\"ur Luft- und Raumfahrt e. V. (DLR), Linder H\"ohe, 51147 K\"oln, Germany\label{aff42}
\and
INFN section of Naples, Via Cinthia 6, 80126, Napoli, Italy\label{aff43}
\and
Institute for Astronomy, University of Hawaii, 2680 Woodlawn Drive, Honolulu, HI 96822, USA\label{aff44}
\and
Dipartimento di Fisica e Astronomia "Augusto Righi" - Alma Mater Studiorum Universit\`a di Bologna, Viale Berti Pichat 6/2, 40127 Bologna, Italy\label{aff45}
\and
Instituto de Astrof\'{\i}sica de Canarias, E-38205 La Laguna, Tenerife, Spain\label{aff46}
\and
Institute for Astronomy, University of Edinburgh, Royal Observatory, Blackford Hill, Edinburgh EH9 3HJ, UK\label{aff47}
\and
Jodrell Bank Centre for Astrophysics, Department of Physics and Astronomy, University of Manchester, Oxford Road, Manchester M13 9PL, UK\label{aff48}
\and
European Space Agency/ESRIN, Largo Galileo Galilei 1, 00044 Frascati, Roma, Italy\label{aff49}
\and
ESAC/ESA, Camino Bajo del Castillo, s/n., Urb. Villafranca del Castillo, 28692 Villanueva de la Ca\~nada, Madrid, Spain\label{aff50}
\and
Universit\'e Claude Bernard Lyon 1, CNRS/IN2P3, IP2I Lyon, UMR 5822, Villeurbanne, F-69100, France\label{aff51}
\and
Institut de Ci\`{e}ncies del Cosmos (ICCUB), Universitat de Barcelona (IEEC-UB), Mart\'{i} i Franqu\`{e}s 1, 08028 Barcelona, Spain\label{aff52}
\and
Instituci\'o Catalana de Recerca i Estudis Avan\c{c}ats (ICREA), Passeig de Llu\'{\i}s Companys 23, 08010 Barcelona, Spain\label{aff53}
\and
Institut de Ciencies de l'Espai (IEEC-CSIC), Campus UAB, Carrer de Can Magrans, s/n Cerdanyola del Vall\'es, 08193 Barcelona, Spain\label{aff54}
\and
UCB Lyon 1, CNRS/IN2P3, IUF, IP2I Lyon, 4 rue Enrico Fermi, 69622 Villeurbanne, France\label{aff55}
\and
Departamento de F\'isica, Faculdade de Ci\^encias, Universidade de Lisboa, Edif\'icio C8, Campo Grande, PT1749-016 Lisboa, Portugal\label{aff56}
\and
Instituto de Astrof\'isica e Ci\^encias do Espa\c{c}o, Faculdade de Ci\^encias, Universidade de Lisboa, Campo Grande, 1749-016 Lisboa, Portugal\label{aff57}
\and
INAF-Istituto di Astrofisica e Planetologia Spaziali, via del Fosso del Cavaliere, 100, 00100 Roma, Italy\label{aff58}
\and
INFN-Padova, Via Marzolo 8, 35131 Padova, Italy\label{aff59}
\and
Aix-Marseille Universit\'e, CNRS/IN2P3, CPPM, Marseille, France\label{aff60}
\and
Space Science Data Center, Italian Space Agency, via del Politecnico snc, 00133 Roma, Italy\label{aff61}
\and
INFN-Bologna, Via Irnerio 46, 40126 Bologna, Italy\label{aff62}
\and
Institute of Theoretical Astrophysics, University of Oslo, P.O. Box 1029 Blindern, 0315 Oslo, Norway\label{aff63}
\and
Jet Propulsion Laboratory, California Institute of Technology, 4800 Oak Grove Drive, Pasadena, CA, 91109, USA\label{aff64}
\and
Department of Physics, Lancaster University, Lancaster, LA1 4YB, UK\label{aff65}
\and
Felix Hormuth Engineering, Goethestr. 17, 69181 Leimen, Germany\label{aff66}
\and
Technical University of Denmark, Elektrovej 327, 2800 Kgs. Lyngby, Denmark\label{aff67}
\and
Cosmic Dawn Center (DAWN), Denmark\label{aff68}
\and
Max-Planck-Institut f\"ur Astronomie, K\"onigstuhl 17, 69117 Heidelberg, Germany\label{aff69}
\and
NASA Goddard Space Flight Center, Greenbelt, MD 20771, USA\label{aff70}
\and
Universit\'e de Gen\`eve, D\'epartement de Physique Th\'eorique and Centre for Astroparticle Physics, 24 quai Ernest-Ansermet, CH-1211 Gen\`eve 4, Switzerland\label{aff71}
\and
Department of Physics, P.O. Box 64, University of Helsinki, 00014 Helsinki, Finland\label{aff72}
\and
Helsinki Institute of Physics, Gustaf H{\"a}llstr{\"o}min katu 2, University of Helsinki, 00014 Helsinki, Finland\label{aff73}
\and
Laboratoire d'etude de l'Univers et des phenomenes eXtremes, Observatoire de Paris, Universit\'e PSL, Sorbonne Universit\'e, CNRS, 92190 Meudon, France\label{aff74}
\and
SKAO, Jodrell Bank, Lower Withington, Macclesfield SK11 9FT, UK\label{aff75}
\and
Centre de Calcul de l'IN2P3/CNRS, 21 avenue Pierre de Coubertin 69627 Villeurbanne Cedex, France\label{aff76}
\and
Dipartimento di Fisica "Aldo Pontremoli", Universit\`a degli Studi di Milano, Via Celoria 16, 20133 Milano, Italy\label{aff77}
\and
INFN-Sezione di Milano, Via Celoria 16, 20133 Milano, Italy\label{aff78}
\and
Universit\"at Bonn, Argelander-Institut f\"ur Astronomie, Auf dem H\"ugel 71, 53121 Bonn, Germany\label{aff79}
\and
INFN-Sezione di Roma, Piazzale Aldo Moro, 2 - c/o Dipartimento di Fisica, Edificio G. Marconi, 00185 Roma, Italy\label{aff80}
\and
Department of Physics, Institute for Computational Cosmology, Durham University, South Road, Durham, DH1 3LE, UK\label{aff81}
\and
Universit\'e Paris Cit\'e, CNRS, Astroparticule et Cosmologie, 75013 Paris, France\label{aff82}
\and
CNRS-UCB International Research Laboratory, Centre Pierre Bin\'etruy, IRL2007, CPB-IN2P3, Berkeley, USA\label{aff83}
\and
University of Applied Sciences and Arts of Northwestern Switzerland, School of Engineering, 5210 Windisch, Switzerland\label{aff84}
\and
Institut d'Astrophysique de Paris, 98bis Boulevard Arago, 75014, Paris, France\label{aff85}
\and
Institut d'Astrophysique de Paris, UMR 7095, CNRS, and Sorbonne Universit\'e, 98 bis boulevard Arago, 75014 Paris, France\label{aff86}
\and
Institute of Physics, Laboratory of Astrophysics, Ecole Polytechnique F\'ed\'erale de Lausanne (EPFL), Observatoire de Sauverny, 1290 Versoix, Switzerland\label{aff87}
\and
Telespazio UK S.L. for European Space Agency (ESA), Camino bajo del Castillo, s/n, Urbanizacion Villafranca del Castillo, Villanueva de la Ca\~nada, 28692 Madrid, Spain\label{aff88}
\and
Institut de F\'{i}sica d'Altes Energies (IFAE), The Barcelona Institute of Science and Technology, Campus UAB, 08193 Bellaterra (Barcelona), Spain\label{aff89}
\and
DARK, Niels Bohr Institute, University of Copenhagen, Jagtvej 155, 2200 Copenhagen, Denmark\label{aff90}
\and
Waterloo Centre for Astrophysics, University of Waterloo, Waterloo, Ontario N2L 3G1, Canada\label{aff91}
\and
Department of Physics and Astronomy, University of Waterloo, Waterloo, Ontario N2L 3G1, Canada\label{aff92}
\and
Perimeter Institute for Theoretical Physics, Waterloo, Ontario N2L 2Y5, Canada\label{aff93}
\and
Universit\'e Paris-Saclay, Universit\'e Paris Cit\'e, CEA, CNRS, AIM, 91191, Gif-sur-Yvette, France\label{aff94}
\and
Centre National d'Etudes Spatiales -- Centre spatial de Toulouse, 18 avenue Edouard Belin, 31401 Toulouse Cedex 9, France\label{aff95}
\and
Institute of Space Science, Str. Atomistilor, nr. 409 M\u{a}gurele, Ilfov, 077125, Romania\label{aff96}
\and
Consejo Superior de Investigaciones Cientificas, Calle Serrano 117, 28006 Madrid, Spain\label{aff97}
\and
Universidad de La Laguna, Dpto. Astrof\'\i sica, E-38206 La Laguna, Tenerife, Spain\label{aff98}
\and
Dipartimento di Fisica e Astronomia "G. Galilei", Universit\`a di Padova, Via Marzolo 8, 35131 Padova, Italy\label{aff99}
\and
Institut f\"ur Theoretische Physik, University of Heidelberg, Philosophenweg 16, 69120 Heidelberg, Germany\label{aff100}
\and
Institut de Recherche en Astrophysique et Plan\'etologie (IRAP), Universit\'e de Toulouse, CNRS, UPS, CNES, 14 Av. Edouard Belin, 31400 Toulouse, France\label{aff101}
\and
Universit\'e St Joseph; Faculty of Sciences, Beirut, Lebanon\label{aff102}
\and
Departamento de F\'isica, FCFM, Universidad de Chile, Blanco Encalada 2008, Santiago, Chile\label{aff103}
\and
Universit\"at Innsbruck, Institut f\"ur Astro- und Teilchenphysik, Technikerstr. 25/8, 6020 Innsbruck, Austria\label{aff104}
\and
Satlantis, University Science Park, Sede Bld 48940, Leioa-Bilbao, Spain\label{aff105}
\and
Infrared Processing and Analysis Center, California Institute of Technology, Pasadena, CA 91125, USA\label{aff106}
\and
Instituto de Astrof\'isica e Ci\^encias do Espa\c{c}o, Faculdade de Ci\^encias, Universidade de Lisboa, Tapada da Ajuda, 1349-018 Lisboa, Portugal\label{aff107}
\and
Department of Physics and Astronomy, University College London, Gower Street, London WC1E 6BT, UK\label{aff108}
\and
Mullard Space Science Laboratory, University College London, Holmbury St Mary, Dorking, Surrey RH5 6NT, UK\label{aff109}
\and
Cosmic Dawn Center (DAWN)\label{aff110}
\and
Niels Bohr Institute, University of Copenhagen, Jagtvej 128, 2200 Copenhagen, Denmark\label{aff111}
\and
Universidad Polit\'ecnica de Cartagena, Departamento de Electr\'onica y Tecnolog\'ia de Computadoras,  Plaza del Hospital 1, 30202 Cartagena, Spain\label{aff112}
\and
Dipartimento di Fisica e Scienze della Terra, Universit\`a degli Studi di Ferrara, Via Giuseppe Saragat 1, 44122 Ferrara, Italy\label{aff113}
\and
Istituto Nazionale di Fisica Nucleare, Sezione di Ferrara, Via Giuseppe Saragat 1, 44122 Ferrara, Italy\label{aff114}
\and
INAF, Istituto di Radioastronomia, Via Piero Gobetti 101, 40129 Bologna, Italy\label{aff115}
\and
Astronomical Observatory of the Autonomous Region of the Aosta Valley (OAVdA), Loc. Lignan 39, I-11020, Nus (Aosta Valley), Italy\label{aff116}
\and
Universit\'e C\^{o}te d'Azur, Observatoire de la C\^{o}te d'Azur, CNRS, Laboratoire Lagrange, Bd de l'Observatoire, CS 34229, 06304 Nice cedex 4, France\label{aff117}
\and
Department of Physics, Oxford University, Keble Road, Oxford OX1 3RH, UK\label{aff118}
\and
Universit\'e PSL, Observatoire de Paris, Sorbonne Universit\'e, CNRS, LERMA, 75014, Paris, France\label{aff119}
\and
Universit\'e Paris-Cit\'e, 5 Rue Thomas Mann, 75013, Paris, France\label{aff120}
\and
Dipartimento di Fisica, Sapienza Universit\`a di Roma, Piazzale Aldo Moro 2, 00185 Roma, Italy\label{aff121}
\and
Aurora Technology for European Space Agency (ESA), Camino bajo del Castillo, s/n, Urbanizacion Villafranca del Castillo, Villanueva de la Ca\~nada, 28692 Madrid, Spain\label{aff122}
\and
ICL, Junia, Universit\'e Catholique de Lille, LITL, 59000 Lille, France\label{aff123}
\and
ICSC - Centro Nazionale di Ricerca in High Performance Computing, Big Data e Quantum Computing, Via Magnanelli 2, Bologna, Italy\label{aff124}
\and
Instituto de F\'isica Te\'orica UAM-CSIC, Campus de Cantoblanco, 28049 Madrid, Spain\label{aff125}
\and
CERCA/ISO, Department of Physics, Case Western Reserve University, 10900 Euclid Avenue, Cleveland, OH 44106, USA\label{aff126}
\and
Technical University of Munich, TUM School of Natural Sciences, Physics Department, James-Franck-Str.~1, 85748 Garching, Germany\label{aff127}
\and
Max-Planck-Institut f\"ur Astrophysik, Karl-Schwarzschild-Str.~1, 85748 Garching, Germany\label{aff128}
\and
Laboratoire Univers et Th\'eorie, Observatoire de Paris, Universit\'e PSL, Universit\'e Paris Cit\'e, CNRS, 92190 Meudon, France\label{aff129}
\and
Departamento de F{\'\i}sica Fundamental. Universidad de Salamanca. Plaza de la Merced s/n. 37008 Salamanca, Spain\label{aff130}
\and
Universit\'e de Strasbourg, CNRS, Observatoire astronomique de Strasbourg, UMR 7550, 67000 Strasbourg, France\label{aff131}
\and
Center for Data-Driven Discovery, Kavli IPMU (WPI), UTIAS, The University of Tokyo, Kashiwa, Chiba 277-8583, Japan\label{aff132}
\and
Dipartimento di Fisica - Sezione di Astronomia, Universit\`a di Trieste, Via Tiepolo 11, 34131 Trieste, Italy\label{aff133}
\and
California Institute of Technology, 1200 E California Blvd, Pasadena, CA 91125, USA\label{aff134}
\and
Department of Physics \& Astronomy, University of California Irvine, Irvine CA 92697, USA\label{aff135}
\and
Departamento F\'isica Aplicada, Universidad Polit\'ecnica de Cartagena, Campus Muralla del Mar, 30202 Cartagena, Murcia, Spain\label{aff136}
\and
Instituto de F\'isica de Cantabria, Edificio Juan Jord\'a, Avenida de los Castros, 39005 Santander, Spain\label{aff137}
\and
INFN, Sezione di Lecce, Via per Arnesano, CP-193, 73100, Lecce, Italy\label{aff138}
\and
Department of Mathematics and Physics E. De Giorgi, University of Salento, Via per Arnesano, CP-I93, 73100, Lecce, Italy\label{aff139}
\and
INAF-Sezione di Lecce, c/o Dipartimento Matematica e Fisica, Via per Arnesano, 73100, Lecce, Italy\label{aff140}
\and
CEA Saclay, DFR/IRFU, Service d'Astrophysique, Bat. 709, 91191 Gif-sur-Yvette, France\label{aff141}
\and
Institute of Cosmology and Gravitation, University of Portsmouth, Portsmouth PO1 3FX, UK\label{aff142}
\and
Department of Computer Science, Aalto University, PO Box 15400, Espoo, FI-00 076, Finland\label{aff143}
\and
Caltech/IPAC, 1200 E. California Blvd., Pasadena, CA 91125, USA\label{aff144}
\and
Ruhr University Bochum, Faculty of Physics and Astronomy, Astronomical Institute (AIRUB), German Centre for Cosmological Lensing (GCCL), 44780 Bochum, Germany\label{aff145}
\and
Department of Physics and Astronomy, Vesilinnantie 5, University of Turku, 20014 Turku, Finland\label{aff146}
\and
Serco for European Space Agency (ESA), Camino bajo del Castillo, s/n, Urbanizacion Villafranca del Castillo, Villanueva de la Ca\~nada, 28692 Madrid, Spain\label{aff147}
\and
ARC Centre of Excellence for Dark Matter Particle Physics, Melbourne, Australia\label{aff148}
\and
Centre for Astrophysics \& Supercomputing, Swinburne University of Technology,  Hawthorn, Victoria 3122, Australia\label{aff149}
\and
Department of Physics and Astronomy, University of the Western Cape, Bellville, Cape Town, 7535, South Africa\label{aff150}
\and
DAMTP, Centre for Mathematical Sciences, Wilberforce Road, Cambridge CB3 0WA, UK\label{aff151}
\and
Kavli Institute for Cosmology Cambridge, Madingley Road, Cambridge, CB3 0HA, UK\label{aff152}
\and
Department of Astrophysics, University of Zurich, Winterthurerstrasse 190, 8057 Zurich, Switzerland\label{aff153}
\and
Department of Physics, Centre for Extragalactic Astronomy, Durham University, South Road, Durham, DH1 3LE, UK\label{aff154}
\and
IRFU, CEA, Universit\'e Paris-Saclay 91191 Gif-sur-Yvette Cedex, France\label{aff155}
\and
Oskar Klein Centre for Cosmoparticle Physics, Department of Physics, Stockholm University, Stockholm, SE-106 91, Sweden\label{aff156}
\and
Astrophysics Group, Blackett Laboratory, Imperial College London, London SW7 2AZ, UK\label{aff157}
\and
Univ. Grenoble Alpes, CNRS, Grenoble INP, LPSC-IN2P3, 53, Avenue des Martyrs, 38000, Grenoble, France\label{aff158}
\and
Centro de Astrof\'{\i}sica da Universidade do Porto, Rua das Estrelas, 4150-762 Porto, Portugal\label{aff159}
\and
Instituto de Astrof\'isica e Ci\^encias do Espa\c{c}o, Universidade do Porto, CAUP, Rua das Estrelas, PT4150-762 Porto, Portugal\label{aff160}
\and
HE Space for European Space Agency (ESA), Camino bajo del Castillo, s/n, Urbanizacion Villafranca del Castillo, Villanueva de la Ca\~nada, 28692 Madrid, Spain\label{aff161}
\and
INAF - Osservatorio Astronomico d'Abruzzo, Via Maggini, 64100, Teramo, Italy\label{aff162}
\and
Theoretical astrophysics, Department of Physics and Astronomy, Uppsala University, Box 516, 751 37 Uppsala, Sweden\label{aff163}
\and
Minnesota Institute for Astrophysics, University of Minnesota, 116 Church St SE, Minneapolis, MN 55455, USA\label{aff164}
\and
Mathematical Institute, University of Leiden, Einsteinweg 55, 2333 CA Leiden, The Netherlands\label{aff165}
\and
School of Physics \& Astronomy, University of Southampton, Highfield Campus, Southampton SO17 1BJ, UK\label{aff166}
\and
Institute of Astronomy, University of Cambridge, Madingley Road, Cambridge CB3 0HA, UK\label{aff167}
\and
Department of Astrophysical Sciences, Peyton Hall, Princeton University, Princeton, NJ 08544, USA\label{aff168}
\and
Space physics and astronomy research unit, University of Oulu, Pentti Kaiteran katu 1, FI-90014 Oulu, Finland\label{aff169}
\and
Center for Computational Astrophysics, Flatiron Institute, 162 5th Avenue, 10010, New York, NY, USA\label{aff170}}    %

%
%
\abstract{
This study explores a rare population of sources in a currently uncharted region of spectroscopic redshift space in the Euclid Quick Data Release (Q1), and is intended potentially to support upcoming spectroscopic studies.
Our goal is to identify and investigate a population of sources characterised by highly ionised emission lines in their spectra, which are indicative of active galactic nucleus activity, extreme shock phenomena, or Wolf--Rayet stars.
A comprehensive visual inspection of spectra is conducted to ensure the reliability of the sample, focusing on the simultaneous detection of both \NeV\ and \OII\ emission-line measurements, a condition that restricts the \Euclid spectroscopic redshift range to $z=2.48$--3.88.
To characterise this population, we analysed the morpho-spectrophotometric properties of their host galaxies. This allowed for a direct comparison with control sources that exhibit similar \OII\ properties and spectroscopic redshifts, but not \NeV\ lines.
We identify sources solely based on spectroscopic criteria in the redshift range beyond the \halpha\ regime.
Encompassing 65 potential \NeV\ candidates, the resulting sample delivers the first systematic probe of these \NeV\ candidate emitters at high redshift. 
We found a good agreement, within 1\,$\sigma$, between the spectral measurements calculated using both direct integration and Gaussian fitting methodologies. The \NeV\ candidates exhibit colours similar to bright QSOs,
with only a few in the tail of very red quasars.
We observed a higher stellar mass content, a lower continuum around the 4000\,\AA\ break, and a similar S\'{e}rsic index distribution compared to the control sample.
This unique sample paves the way for a wide range of scientific investigations, which will be pursued in the forthcoming data releases.}


\titlerunning{Highly ionised emitters with \Euclid spectroscopy}
\authorrunning{Euclid Collaboration: D.~Vergani et al.}
   
\maketitle
%
%
%
%
\section{\label{sc:Intro}Introduction}

Understanding the full diversity of astrophysical phenomena is crucial for unravelling the complex tapestry of galaxy evolution \citep[e.g.,][]{Silk1998, SomNao2017}. However, conventional multi-wavelength diagnostics often miss rare and elusive populations, particularly those associated with extreme energy release and rapid evolutionary changes in their host galaxies \citep{Kennicutt2012}. Active galactic nuclei (AGN) and extreme  shock phenomena represent key manifestations of such  activity, characterised by the release of vast amounts of energy and possessing the potential to significantly regulate star formation through feedback processes \citep{Fabian2012, King2015, Harrison2018}.

These highly energetic processes involve the production of characteristic emission lines from species with substantial ionisation potentials. Among these, the \NeV\ emission line, originating from quadruply ionised neon requiring an ionisation potential of \SI{97.1}{eV}, serves as a powerful diagnostic tool for identifying targets with exceptionally high ionisation levels. While primarily attributed to the intense radiation fields within the inner regions of AGN \citep{Cleri2023, PereiraSantaella2010,Chisholm2024}, \NeV\ emission can also be indicative of extraordinary shock phenomena in extremely intense star-formation events, or the presence of Wolf--Rayet stars \citep{Zeimann2015}. Recent studies underscore the value of \NeV\ emitters for AGN selection, especially for detecting highly obscured and Compton-thick AGN that often elude traditional X-ray or infrared surveys \citep{Gilli2007, Lansbury2017, HickoxAlexander2018, Mazzolari2024, Barchiesi2024}. Given their frequent presence in massive, star-forming galaxies \citep{Vergani2018}, investigating these \nevno-selected sources is essential for a more thorough understanding of AGN and their complex role in galaxy evolution.

The \Euclid mission provides deep near-infrared photometry at high spatial resolution, alongside slitless spectroscopy for measuring redshifts and emission lines \citep{EuclidSkyOverview,Q1-TP006,Q1-TP007}. 
The unique combination of area, depth, and ancillary data offers an unprecedented opportunity to select purely through spectral diagnostics a \NeV\ candidate sample at $z=2.48$--3.88, a range where the key diagnostic lines fall within the \Euclid red grism.
We present the first systematic identification and characterisation of this rare population using initial data from the \cite{Q1cite} providing crucial insights into their host galaxy properties and, in the future, their role in regulating star formation.
  
\section{Datasets}
\label{sc:data}

A comprehensive overview of the Q1 data release is provided in \citet{Q1-TP001}, \citet{Q1-TP002}, \citet{Q1-TP003}, and \citet{Q1-TP004}. A concise summary of the mission’s scientific goals is available in \citet{EuclidSkyOverview}. Briefly, \Euclid surveyed approximately $63\,\mathrm{deg}^2$ of the extragalactic sky for Q1, covering three distinct regions and sources from the Euclid Deep Field North (EDF-N), Euclid Deep Field South (EDF-S), and Euclid Deep Field Fornax (EDF-F).
Observations were carried out in one visible band \citep[$\IE$,][]{EuclidSkyVIS}, three {NIR} photometric bands \citep[$\YE$, $\JE$, and $\HE$,][]{EuclidSkyNISP}, and using the red grism, resulting in two-dimensional spectra from the Near Infrared Spectrometer and Photometer (NISP) reduced using the dedicated Spectroscopic Image Reduction (SIR) processing function \citep[][]{Q1-TP006}. The spectroscopy (SPE) processing function within the \Euclid\ pipeline automatically analyses one-dimensional spectra to determine redshifts, line fluxes, and spectral classifications \citep[][]{Q1-TP007}.

This analysis used the newly available Q1 data set combined with observations at $3.6\,\micron$ and $4.5\,\micron$ with the Infrared Array Camera \citep[IRAC,][]{Fazio2004} on {\it Spitzer} \citep{Werner2004} as compiled by \cite{Q1-SP031}. 

\begin{figure}[h!]     
    \includegraphics[angle=0,width=1.\hsize]{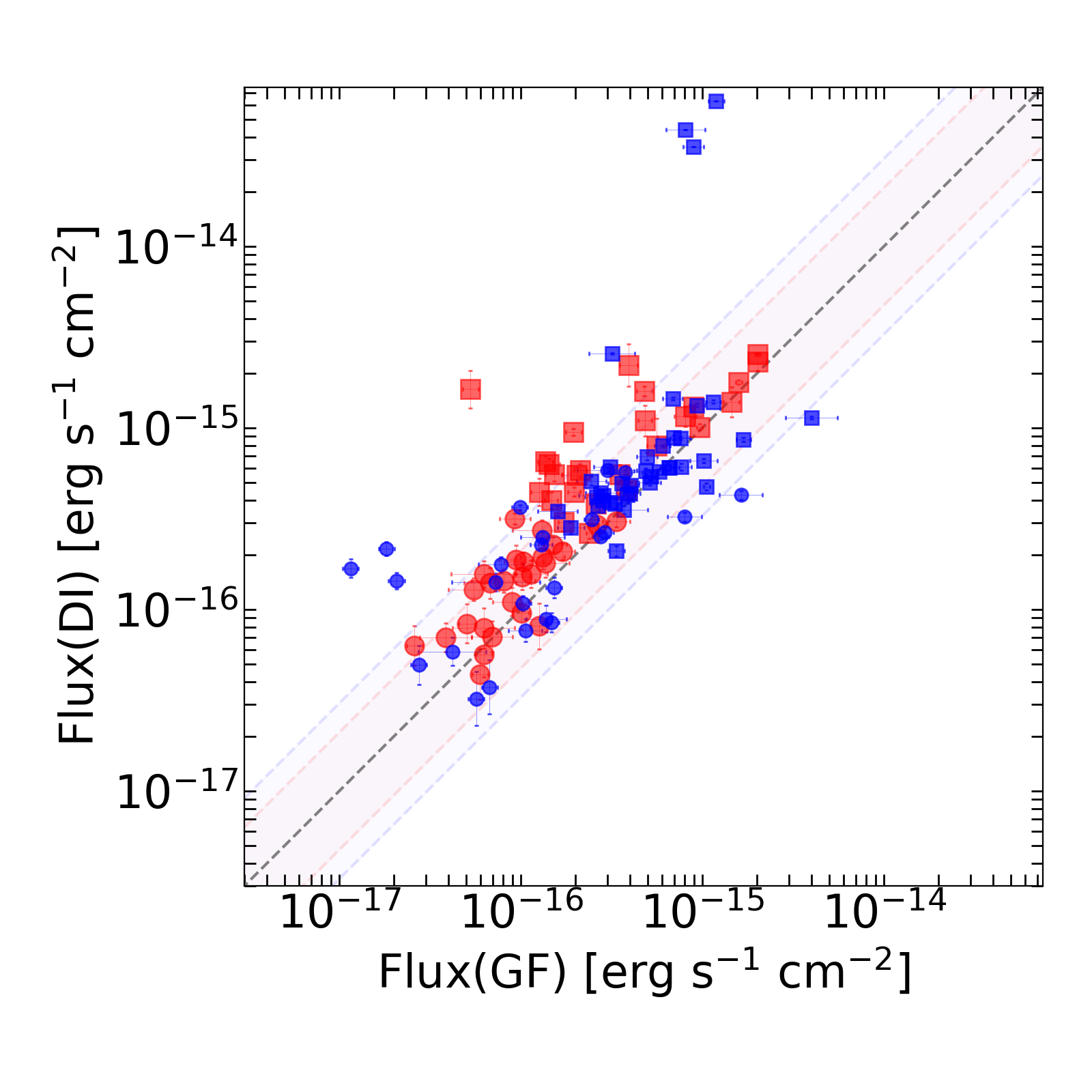}
        \caption{Comparison of the fluxes measured via the Gaussian fit (GF, $x$-axis) and the direct integration (DI, $y$-axis) for the \NeV\ and \OII\ emission lines used in this analysis (squares and circles, respectively). QSO- and galaxy-classified \NeV\ emitters are plotted with red and blue symbols, respectively. The dashed line represents the $x=y$ relation. The bands represent the $\pm$1\,$\sigma$ distributions of each category.}
    \label{fig:gfdi}
\end{figure}

\subsection{Sample selection}

Here we describe the conditions applied to select the samples based on the Q1 products available from the ESA Science Analysis System (SAS\footnote{The SAS is available at \url{https://eas.esac.esa.int/}.}).
In all samples used in this work we selected the most reliable \Euclid spectroscopic redshift (or {\small\texttt{spe\_z}} with {\small\texttt{spe\_rank=0}}) among the five available solutions in 
{\small\texttt{catalogue.spectro\_zcatalog\_spe\_qso\_candidates}} and {\small\texttt{catalogue.spectro\_zcatalog\_spe\_galaxy\_candidates}},
based on spectroscopic classification derived from template fitting (or {\small\texttt{spe\_class}} in {\small\texttt{spectro\_zcatalog\_spe\_classification}}). 
This classification utilizes composite templates derived from major surveys: galaxy spectra are based on a combination of six Bruzual--Charlot \citep[BC03][]{BC03} continua and 14 VVDS-based emission-line ratio templates (i.e., VIMOS VLT Deep Survey, \citealp{VVDS2013}); quasars use a single mean SDSS spectrum  \citep{VandenBerk2001}; and stars are modeled by 36 stellar spectra from the European Southern Observatory. We refer for a full description to Sect.\,2.1 of \cite{Q1-TP007}.
In the Q1 release, all spectral line properties (such as flux, signal-to-noise ratio, and central wavelength in {\small\texttt{spectro\_line\_features\_catalog\_spe\_line\_features\_cat}} and continuum indices in {\small\texttt{spectro\_line\_features\_catalog\_spe\\\_continuum\_features\_cat}}) were computed assuming that all observed line measurements correspond to a galaxy redshift solution\footnote{This limitation stems from a known bug in the data processing pipeline and is scheduled for correction in subsequent releases.}, as detailed in Sect.\,5 of \cite{Q1-TP007}.

To address this limitation, we measured for {\small\texttt{spe\_class=qso}} the line features at the corresponding spectroscopic redshifts.
The sources in the three fields (EDF-N, EDF-F, and EDF-S) were analysed together, without separating them by field.

To select the target sample of the \NeV\ candidate emitters, in addition to selecting the most probable spectroscopic solution from the five available options, we required the simultaneous presence of the \OII\ and \NeV\ emission-line measurements obtained from two distinct methodologies: direct integration (DI) and Gaussian fitting (GF). 
In DI all the pixels of the emission line above the continuum are integrated; in GF we derived a single Gaussian model on the lines considered in this work \citep{Q1-TP007}.
The requirement that both methodologies converge to a consistent result, as observed in Fig.\,\ref{fig:gfdi}, further strengthens the reliability of the sample. Applying these criteria, we obtained an initial sample of 116 sources, comprising 72 galaxy-classified targets and 44 QSO-classified targets.

We then applied rigorous and systematic visual inspection of each spectrum facilitated by the SQ tool (\texttt{SpectraPyle}; \cite{EP-Quai}).
In particular, we excluded spectra affected by systemic factors or noise features near the wavelengths of interest, line profiles inconsistent with physically plausible emission (e.g., unphysical asymmetries or widths, and spurious features), anomalous continuum shapes, residuals from sky subtraction, flux discontinuities, or improper background removal.
Exclusions made during the visual inspection phase were due to data quality issues of instrumental origin.
These quality checks were designed to maximise the reliability of the final candidate sample used in the analysis because the explored redshift regime falls outside the typical H$\alpha$ range and conventional targeting strategies exhibit reduced efficacy.
These criteria yielded a sample of 65 sources from the initial 116 sources, comprising 39 \NeV\ galaxy-classified targets and 26 \NeV\ QSO-classified target sources, spanning a redshift range of $z=2.48$--3.88, a range where the target diagnostic lines fall within the \Euclid grism setup's wavelength coverage.

We generated a control catalogue of 190 sources selected by applying the same criteria used for the \NeV\ target sample, except for the specific \NeV\ constraint.  Instead, we required \Euclid photometric redshifts to be consistent within 10\%. This requirement is considered necessary because spectroscopic redshift estimates for most of these sources rely primarily on a single emission line, and agreement with photometric redshifts helps reduce the risk of catastrophic failures. 

From this control catalogue, we constructed a matched control sample that is comparable in size and optimized to reproduce the target sample's distributions in both redshift and \OII\ flux. We refer to this data set as the  control sample.
The dispersion is assessed both through the $\text{Interquartile Range}$ ($\text{IQR}$) of the control catalogue of 190 sources and via bootstrap resampling, with the $\text{min-max}$ band of the empirical cumulative distribution function quantifying the full variability across the generated realizations.

The spectroscopic control sample was subjected to further refinement owing to the previously mentioned challenges associated with spectroscopy and the absence of a strict requirement for the presence of multiple emission lines, thus relying solely on concordance with photometric redshifts. Each spectrum of the 190 sources of the control sample was meticulously visually inspected for spurious features, resulting in a final high-quality spectroscopic control sample of 50 sources.

\begin{figure}[h!]
    \centering
\includegraphics[width=.5\textwidth]%
{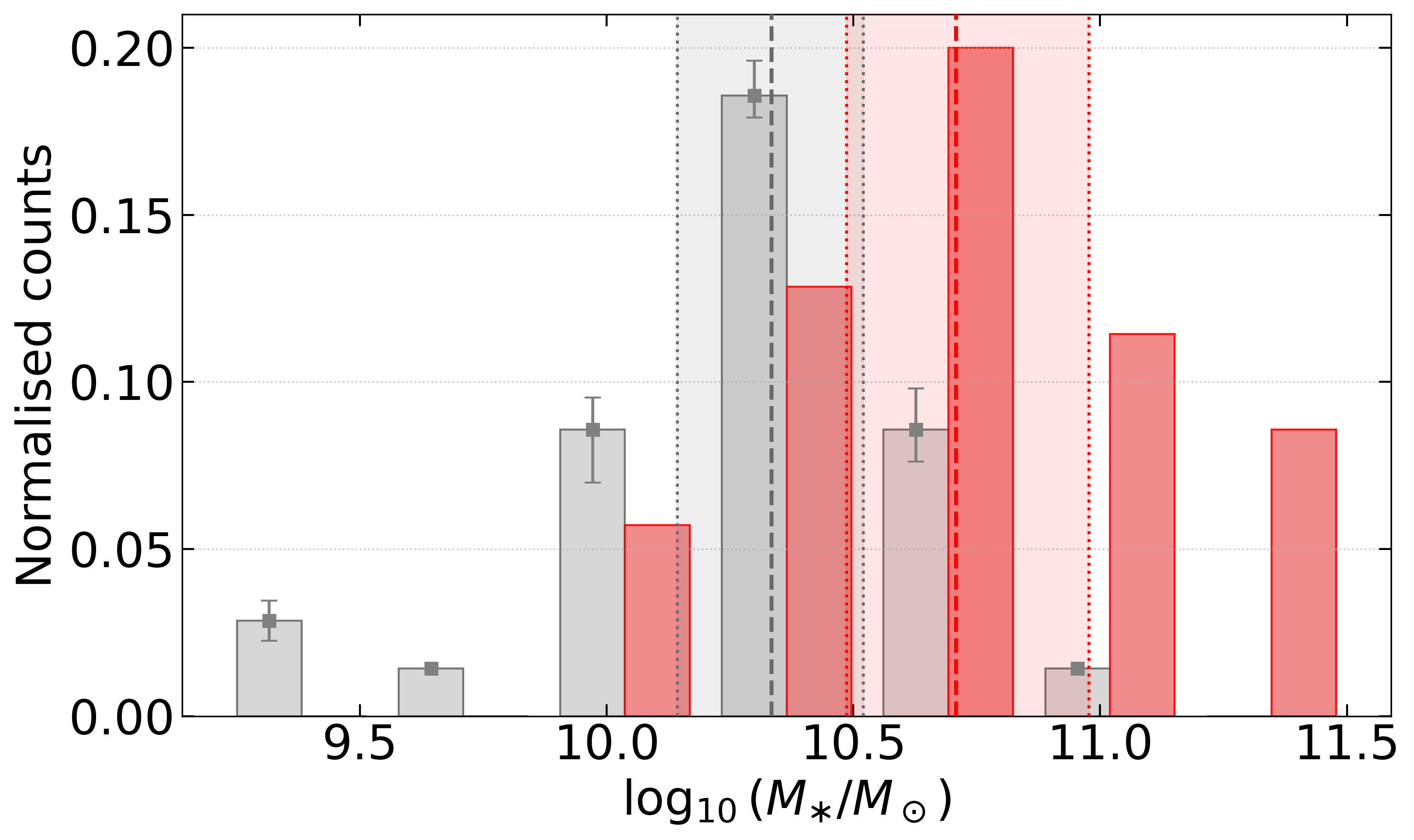}
\includegraphics[width=.5\textwidth]{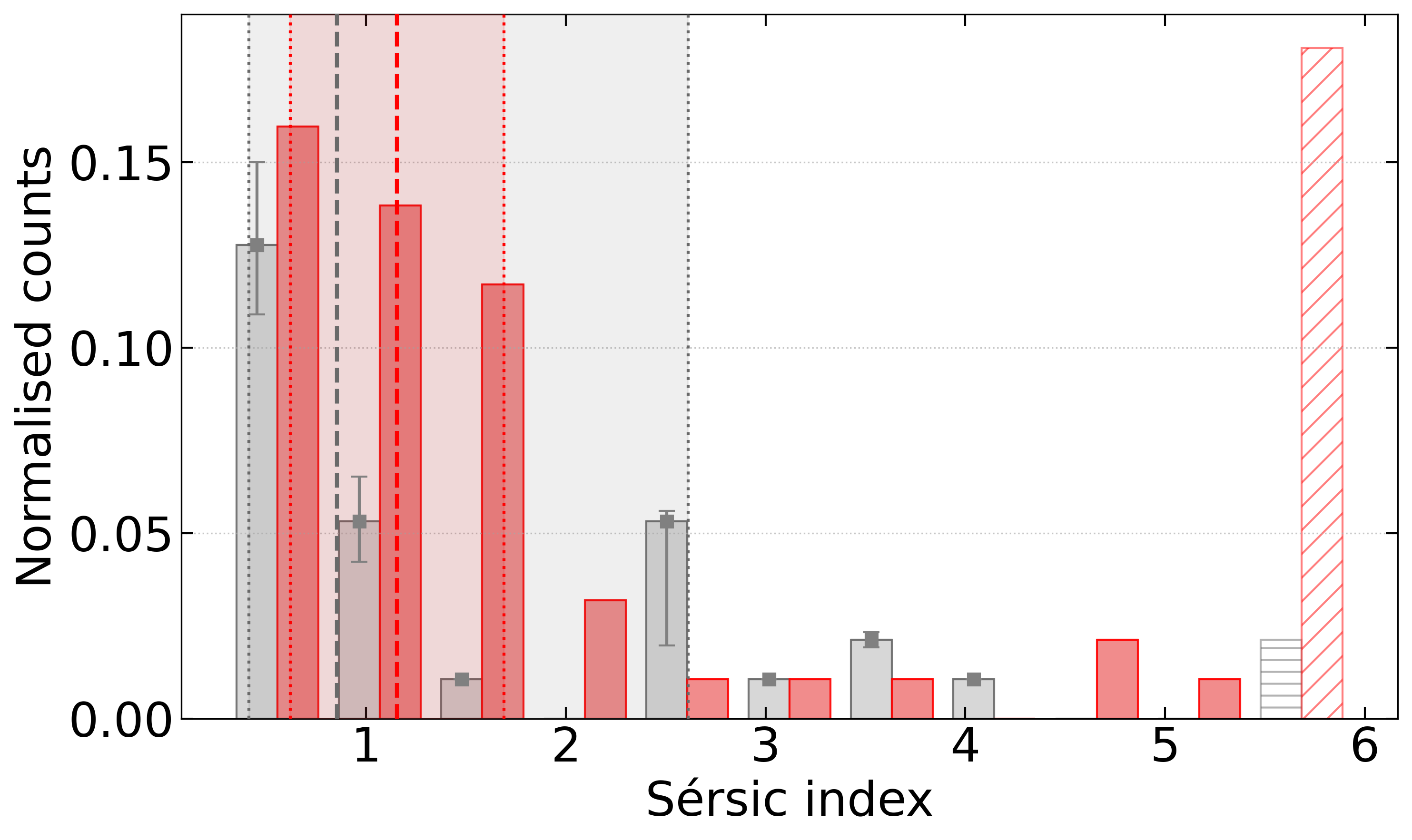}
        \caption{Distributions of the median stellar mass and the S\'{e}rsic index for the objects in the target sample (red) and control sample (grey). The dashed lines indicate the median value, while the vertical dotted lines and shaded bands represent the interquartile ranges(IQRs) of the distributions.  
        The error bars represent the maximum dispersion in the parameters estimated via the bootstrap resampling and the IQR. The hatched bars represent the sources excluded from the analysis.}
    \label{fig:enia}
\end{figure}

\section{Results and discussion}\label{sc:results}

\subsection{Sample Multi-Wavelength Characterization}\label{ss:characterization}

The spectroscopic redshifts of the 65 \NeV\ candidate emitters were cross-verified with their \Euclid photometric redshifts \citep{Q1-TP005} and external spectroscopic surveys (C. Saulder, M. Mezcua, M. Siudek priv. comm.).
However, at high redshifts, the spectroscopic samples are still very scarce, while for faint objects with limited photometry, classification and photometric redshifts are less reliable. Probably due to a combination of these two factors, we did not find concordance in the photometric redshift catalogue, nor were there matches in the compilation of spectroscopic redshifts.

This mismatch is particularly noteworthy for Data Release 1 of the Dark Energy Spectroscopic Instrument (DR1-DESI), where, despite its promising statistical abundance of sources, the specific targeting strategies trace distinct populations across redshift \citep{DESI2025}.
We searched for radio counterparts for all 65 \NeV\ candidates in Low Frequency Array (LOFAR)  data, specifically observations at 144\,MHz with a typical angular resolution of approximately 6\arcsecond\ \citep{Bondi2024, Q1-SP011}. While a possible radio detection exists for a single candidate, the presence of a nearby low-redshift source within the LOFAR beam precludes a definitive association of the radio emission with this \NeV\ candidate emitter.

None of the 65 \NeV\ candidate emitters are associated with an X-ray emission, when we cross-matched with the available \Euclid catalogues of counterparts to X-ray sources \citep{Q1-SP003}. This can be due to multiple reasons, including the small sample size combined with the shallowness and/or incomplete coverage of the X-ray data available for most of the three fields.
The lack of X-ray counterparts in our sample is consistent with the findings of \cite{Barchiesi2024}, who analysed a sample of 94 \NeV-selected type 2 AGN in the COSMOS field at $z=0.6$--1.2, and found that over two-thirds of the sources are heavily obscured, with approximately 20\% identified as candidate Compton-thick AGN. AGN detected via the \NeV\ emission line are rarely identified using conventional AGN‑selection methods based on shallow X‑ray observations, mid-IR colour criteria, or standard emission‑line diagnostic diagrams \citep{Mignoli2013}, potentially indicating a distinct evolutionary phase \citep[{and others}]{Vergani2018}.

\cite{Q1-SP003} used two machine-learning algorithms, which based solely on \Euclid photometry, could provide the probability of a source being extragalactic (${P}_{\textrm{Gal}}$), and being an X-ray emitter (${P}_{\textrm{X-ray}}$). 
The same algorithms have been applied to the \NeV\ and the control samples of 190 sources. 
With the exception of two sources in the control sample and seven sources in the \NeV\ sample, they all have probability >60\% of being extragalactic. 
However, only 4\% and 10\% of the parent and target sample sources, respectively, have a greater than 5\% probability of being X-ray emitters.
A comparison of the ${P}_{\textrm{X-ray}}$ parameter between the \NeV\ and the control sample indicated a statistically significant difference in their X-ray probability distributions. This distinction is supported by both Kolmogorov--Smirnov test ($p=0.0007$) and Mann–Whitney U ($U=8044.0$, $p=0.0003$) tests.
However, \cite{Q1-SP003} warn that these probabilities inherently carry an approximate 10\% incompleteness (false negatives) for X-ray source identification.
Furthermore, the limited size of its training sample may reduce the reliability of the Random Forest algorithm when classifying rare or underrepresented objects, which may contribute to the lack of X-ray identification for our candidates.

\begin{figure}
    \centering
    \includegraphics[angle=0,width=1.0\hsize]{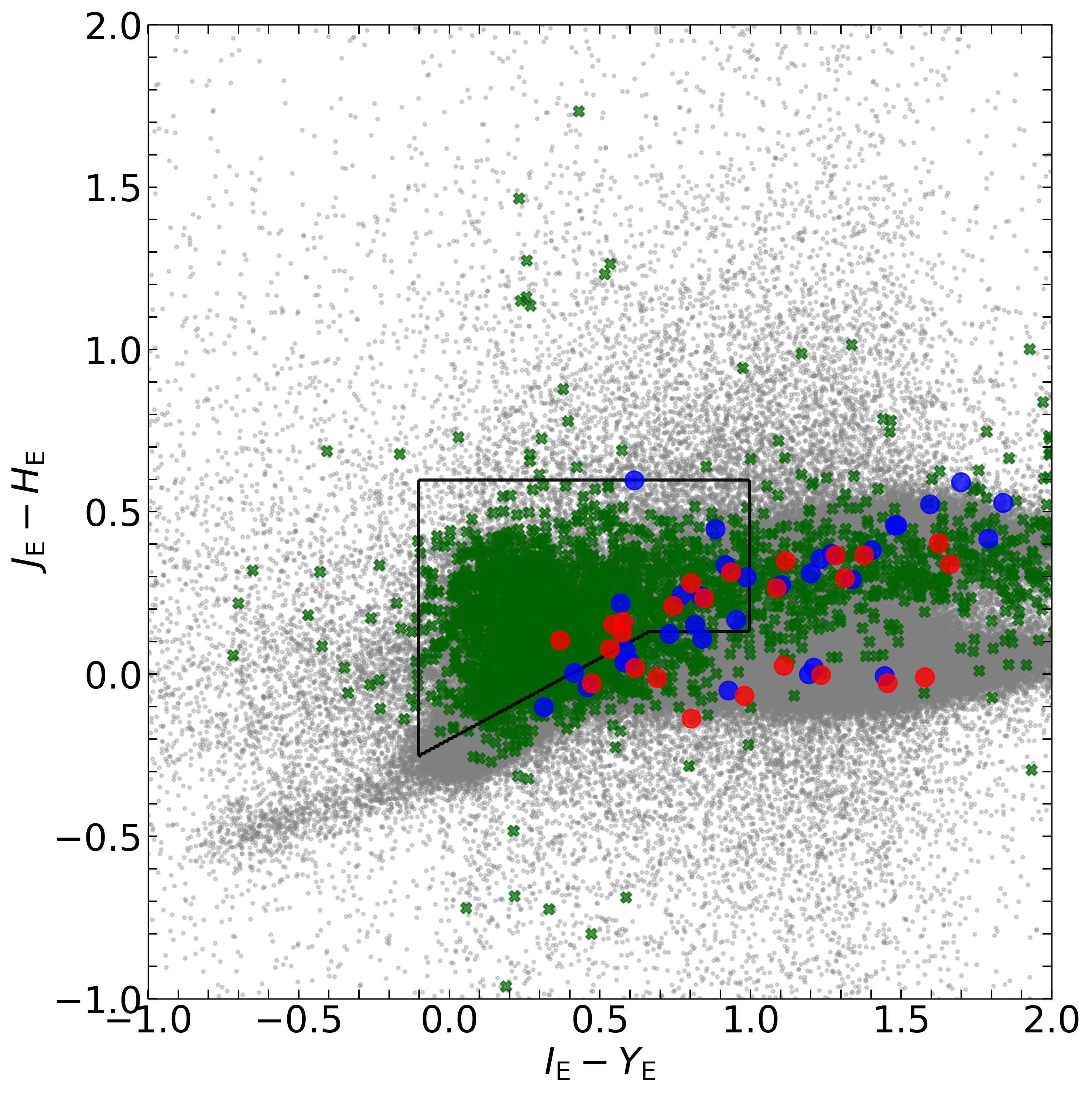} 
    \includegraphics[angle=0,width=0.95\hsize]{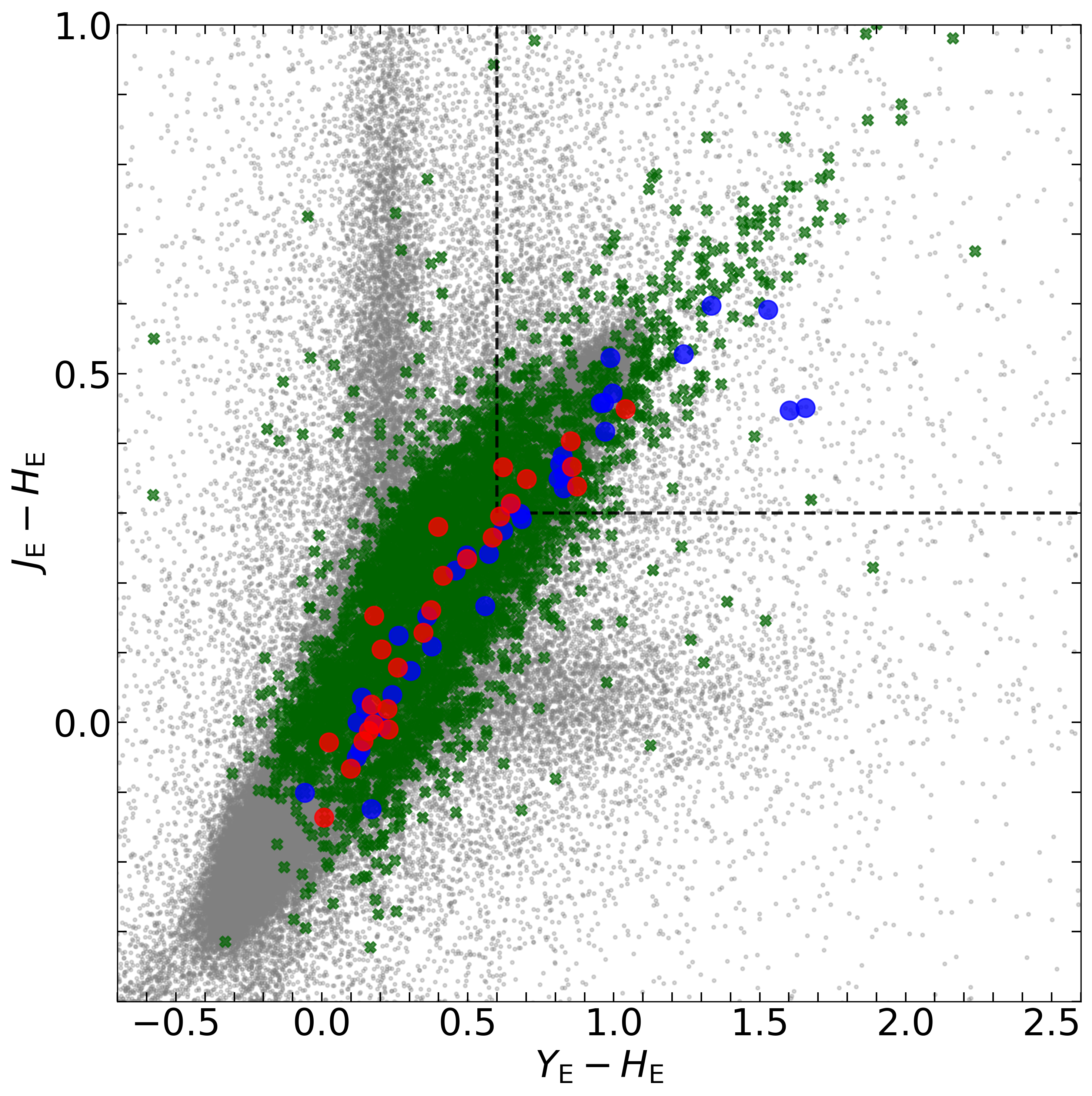}
    \caption{{\it Top}: $(\IE - \YE)$ versus $(\JE - \HE)$ diagram showing the  \NeV\ galaxy-classified targets and \NeV\ QSO-classified target sources  (in blue and red circles) plotted with the \Euclid spectroscopically identified quasars in F25 (green). We overplot the region proposed in \citetalias{Q1-SP027} that identifies the QSO candidates (black-solid line), with the stellar locus found below this boundary at approximately \JE$-$\HE$< 0.1$ represented by purple contours in their Fig.\,6. The grey points are the compact \Euclid\ sources in \citetalias{Q1-SP027}. {\it  Bottom}: $(\YE - \HE)$ versus $(\JE - \HE)$ diagram with the region proposed in \citetalias{Q1-SP023}  that identifies the red QSO (black-dashed line). Symbols are as in the upper panel.}
    \label{fig:jhiy}
\end{figure}

Figure\,\ref{fig:enia} illustrates the distribution of key physical properties for the objects in our samples, specifically showing histograms for the stellar mass and the S\'{e}rsic index (\nSersic). We used the median stellar masses derived by \cite{Q1-SP031} complementing the \Euclid data with public IRAC observations computed at the spectroscopic \Euclid redshifts of the sources.
The \NeV\ candidates exhibit a statistically significant larger stellar mass $\ \logten(M_*/M_\odot)_{\text{\nevno}}=10.71$ (inter quartile range, or IQR: 0.49 representing the central 50\% of the distribution) with respect to the {control samples}, $ \logten(M_*/M_\odot)_{\text{control}}=10.33$ (IQR: 0.38). The Kolmogorov--Smirnov test result confirms the significant difference ($p=2.365 \times 10^{-4}$).

\begin{figure}
    \centering
    \includegraphics[width=.5\textwidth]{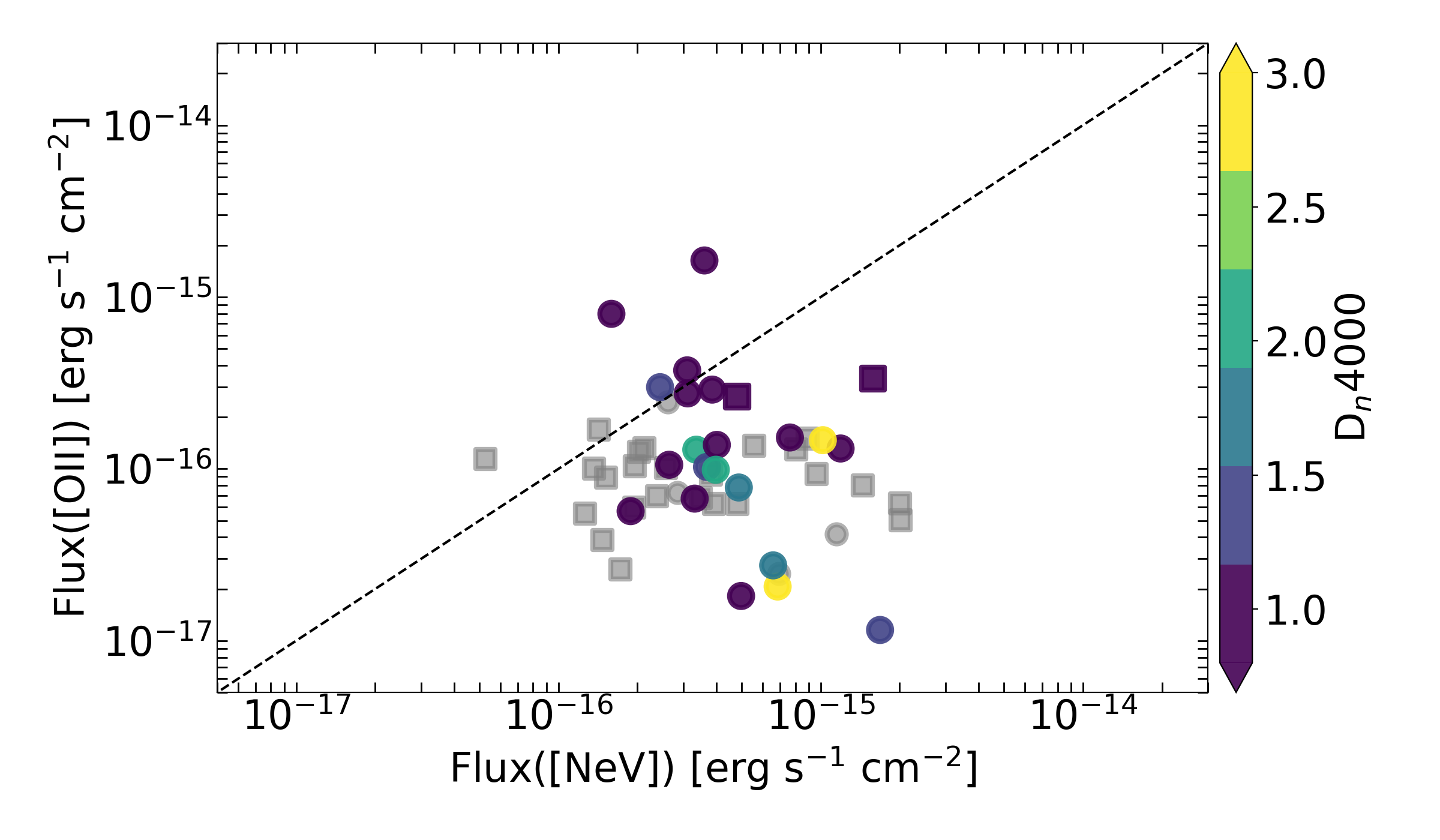}
    \caption{Distribution of the \NeV- ($x$-axis) and \OII-flux ($y$-axis) for the candidate emitters as a function of the  4000\,\AA\ break ({D$_{\rm n}$4000}, in colours when measured, in grey otherwise). Symbols are as in Fig.\,\ref{fig:gfdi}.}
    \label{fig:nevo2}
\end{figure}

We excluded from the morphological analysis presented in the bottom panel of Fig.\,\ref{fig:enia}  those 17 target sources over the 65 with the largest S\'{e}rsic index ($>5.45$) indicative of a potentially lower fit confidence \citep{Quilley25}, and 1 from the control sample.
Among these 17 sources with largest S\'{e}rsic index, 14 sources exhibit morphological properties characteristic of a high probability of being a point-like source, or \texttt{MUMAX\_MINUS\_MAG} $< -2.6$ \citep{Q1-TP004}. 

The S\'{e}rsic index distributions between the \NeV\ candidates and the control sample do not show a statistically significant difference based on the Kolmogorov--Smirnov test ($D=0.2431$, $p=0.2158$). Despite this, the target sample exhibits a slightly lower median S\'{e}rsic index and a moderately tighter Interquartile Range (IQR), with \nSersic$_{\text{, NeV}} = 1.15$ (IQR: $1.07$) compared to \nSersic$_{\text{, control}}=0.85$ (IQR: $2.20$); the visual morphology suggests the \NeV\ candidates are generally slightly concentrated towards lower, disk-like values. Nevertheless, a small but significant number of sources with genuinely non-exponential light profiles (\nSersic$>2.00$) persist (6/65).
The \NeV\ galaxy-classified and \NeV\ QSO-classified sources show a similar distribution, without any segregation toward high or low values. 

In Fig.\,\ref{fig:jhiy}, we show the distribution of highly ionised line sources in the \Euclid\ colour-colour space. Our sample is compared with the spectroscopically confirmed bright QSO sample from Fu et al. (in prep., hereafter F25). 
We overplot the two regions proposed by \citet[][hereafter \citetalias{Q1-SP027}]{Q1-SP027} to select compact AGN and with the red QSO locus identified by \citet[][hereafter \citetalias{Q1-SP023}]{Q1-SP023}.  

The top panel of this figure shows that target \NeV\ candidates are overlapping with the redder \IE$-$\YE\ tail of the spectroscopic bright QSOs (F25), and are not confined to the QSO region defined by \citetalias{Q1-SP027} in the \IE$-$\YE\ versus \JE$-$\HE\ (\JE\HE\_\IE\YE) diagram, but rather extend over a broader area. However, our candidates do not exhibit the distinct colour properties of red quasars -- namely, high \YE$-$\HE\ ($> 0.7$) and \JE$-$\HE\  ($> 0.3$) values -- as required by the \citetalias{Q1-SP023}  criteria (dashed box in the bottom panel of Fig.\,\ref{fig:jhiy}).
It is noteworthy that the region occupied by the stellar locus as identified in M25 (below the black-solid region in figure) is populated by \NeV\ candidates exhibiting both high point-like probability ($\texttt{MUMAX\_MINUS\_MAG} < -2.6$) and large S\'{e}rsic indices (\nSersic$> 5.45$). The nature of these objects, classified morphologically as point-like, will be subject to further investigation to confirm their likely stellar identity.

\begin{table*}
\caption{Median Physical Properties of the \NeV\ candidates and \OII\ emitters}
\label{tab:sample_properties}
\centering
\begin{tabular}{lccc}
\hline\hline
Physical property & Unit & \NeV\ candidates (target) & \OII\ emitters (parent) \\
& & (Median [IQR]) & (Median [IQR]) \\\hline
Spectroscopic redshift & --- & $2.8278 [2.6717, 2.9576]$ & $2.6545 [2.5397, 2.8273]$  \\
Stellar mass ($\log_{10}(M_*/M_{\odot})$) & --- & $10.71$ [10.49, 10.98] & $10.33$ [10.14, 10.52] \\
S\'{e}rsic index (\nSersic) & --- & $1.15$ [0.62, 1.69] & $0.85$ [0.41, 2.61] \\
{D$_{\rm n}$4000} index & --- & $0.98 \pm 0.02$ & $1.06 \pm 0.03$ \\
Flux \NeV\ & erg\,s$^{-1}$\,cm$^{-2}$ & ${5.9}\times 10^{-16}$ & <${0.9}\times 10^{-16}$ \\
Flux \OII\ & erg\,s$^{-1}$\,cm$^{-2}$ & ${3.1}\times 10^{-16}$ & ${5.2}\times 10^{-16}$ \\
\hline
\end{tabular}
\end{table*}

\subsection{Spectroscopic Properties}\label{ss:specprop}

Figure~\ref{fig:nevo2} presents the distribution of \NeV\ and \OII\ fluxes, as computed using the GF method described in Sect.\,{2.1}. The data points are represented by coloured symbols, which correspond to the 4000\,\AA\ break (or D$_{\rm n}$4000), used as a proxy for the age of the underlying stellar population \citep{Kauffmann2004}. In this context, lower {D$_{\rm n}$4000} values are indicative, to a first approximation, of younger stellar ages.
Despite the limited sample size, sources exhibiting higher values for the {D$_{\rm n}$4000} index are exclusively those characterised by low \OII\ flux and high \NeV\ flux. Unfortunately, spectroscopic measurements on absorption lines and indices are available for only a fraction of the sample (23/65), caused by a low signal-to-noise ratio, which precludes further detailed analysis or the drawing of definitive conclusions from individual measurements at this time.

\begin{figure}
    \centering
\includegraphics[width=0.5\textwidth]{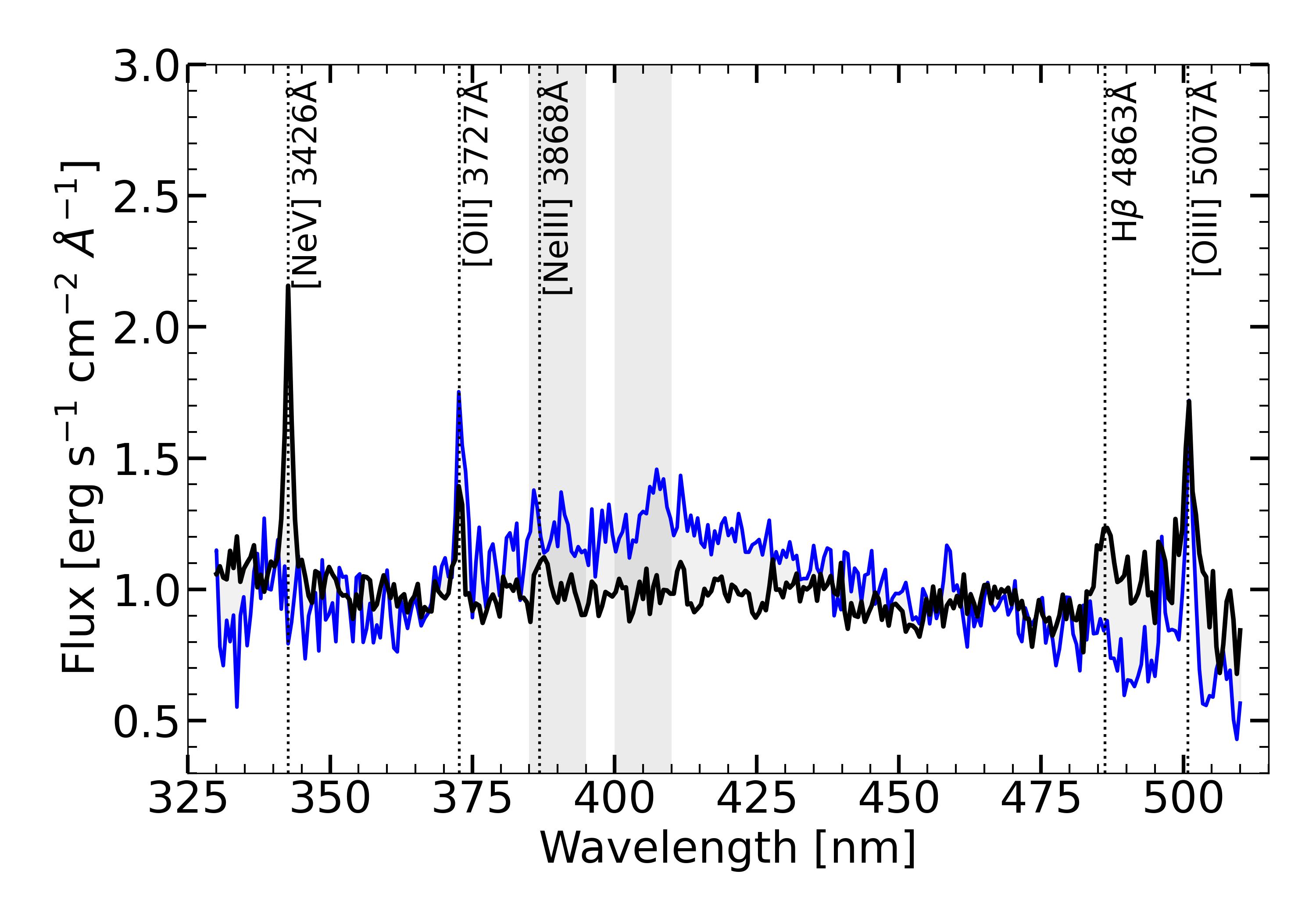}
\caption{Rest-frame smoothed spectrum (3-pixel window) of [Ne\,\textsc{v}] candidate emitters, shown in black, compared with that of the high-quality spectroscopic control sample of 50 \OII\ candidates in blue. The grey shaded bands delineate the wavelength intervals of the continuum used to calculate the 4000\,\AA\ break index.}
    \label{fig:stack}
\end{figure}

We therefore combine the spectra of \NeV\ candidate emitters, and compare with the high-quality spectroscopic control sample of \OII\ emitters (Fig. \ref{fig:stack}).
Individual spectra are first aligned to a common reference redshift; they are then normalised using median scaling to preserve flux conservation, and subsequently combined using the median estimator. For a comprehensive explanation on the spectral stacking technique, we refer the reader to the forthcoming publication of \cite{EP-Quai}.
A 3-pixel moving average filter was applied to the spectra to smooth the data, ensuring flux conservation in the process. 

Emission line fluxes for \NeV\ and \OII\ candidate emitters were derived through a two-step process to ensure robust continuum subtraction in the combined spectra. First, a global continuum was fitted across the wavelength range of 
\SI{320}\nm\ to \SI{460}\nm\ (covered by the majority of the spectra), avoiding known strong emission/absorption features (i.e., lines and the 4000\,\AA\ break region). A second-degree polynomial was employed for this global continuum fit. 
Subsequently, this fitted global continuum was subtracted from the entire original spectrum. The fluxes of individual emission lines were then measured by fitting a single Gaussian profile to the corresponding features in the continuum-subtracted spectrum. 
The 4000\,\AA\ break was calculated following  \citet{Balogh1999}. 
Errors on this index were propagated assuming the RMS of the global continuum residuals as the typical uncertainty in the mean fluxes within these bands.

Comparing the two combined spectra, the highly ionised \NeV\ line is notably absent in the \OII\ candidate emitters, with an upper limit of $0.9 \times 10^{-16}$~erg\,s$^{-1}$\,cm$^{-2}$, or 3\,$\sigma$ RMS. This finding is consistent with the extreme rarity of this line and indicates less extreme ionisation conditions in these galaxies. Conversely, the \NeV\ line is prominently detected in the target sample, exhibiting a flux of ${5.9}\times 10^{-16}$~erg\,s$^{-1}$\,cm$^{-2}$. This stark contrast suggests that the dominant energy sources and ionisation mechanisms differ significantly between the two samples.
The corresponding \OII\ flux exhibits a value of ${5.2} \times 10^{-16} \text{ erg s}^{-1} \text{ cm}^{-2}$ in \OIIno\ emitters, dropping to ${3.1} \times 10^{-16} \text{ erg s}^{-1} \text{ cm}^{-2}$ in the \nevno\ candidates.
Furthermore, when comparing these two distinct combined spectra, we observe a $2.2\,\sigma$ statistically differences in the continuum around the {D$_{\rm n}$4000} break. The \NeV\ candidate sources exhibit a {D$_{\rm n}$4000} index of $0.98 \pm 0.02$, while the \OII\ candidate sample shows a value of $1.06 \pm 0.03$. This disparity in the {D$_{\rm n}$4000} index, a well-established proxy for stellar population age and recent star formation, may indicate a difference in the underlying stellar populations and star-formation histories between the two data sets.

While our analysis focuses on the prominent \NeV\ and \OII\ lines and the 4000\,\AA\ break, we acknowledge other emission features such as the H$\beta$ and [\ion{O}{iii}]$\lambda$4959,5007 complex. These lines, however, fall near the edges of the stacked spectral range and are comprised of a significantly smaller number of individual galaxy spectra. This results in a considerably lower signal-to-noise ratio in these regions. Therefore, we excluded these peripheral lines from our quantitative analysis.

\section{Summary and conclusions}
\label{sc:conclusion}

This study identifies and characterises the first systematic sample of 65 high-redshift ($2.5<z<3.8$) \NeV\ emitting sources within \Euclid's Q1 data set, opening a new window into the properties of these rare objects beyond the H$\alpha$ regime (Tab.\,\ref{tab:sample_properties}). Emerging differences in the stellar mass are already evident in the host galaxy properties of these \NeV\ candidates compared to a control sample matched in \OII\ line properties and redshift, with a 4000\,\AA\ continuum distinct from that of the \OII\ emitter population selected in the same redshift range.
This analysis reveals that while \NeV\ candidate emitters occupy a similar colours to bright QSO (F25), they do not exhibit the same extreme reddening observed in the \citetalias{Q1-SP023} sample.

This high-$z$ sample, combined with our rigorous selection and visual inspection, offers a promising framework for future spectroscopic studies of the phenomena producing these highly ionised lines and their connection to star formation in the host galaxies.

\begin{acknowledgements}
\AckEC 
\AckQone
\AckDatalabs
The authors acknowledge support from the ELSA project. "ELSA: Euclid Legacy Science Advanced analysis tools" (Grant Agreement no. 101135203) is funded by the European Union. 
FR acknowledges the support from the INAF Large Grant "AGN \& \Euclid: a close entanglement" Ob. Fu. 01.05.23.01.14.
Views and opinions expressed are however those of the author(s) only and do not necessarily reflect those of the European Union or Innovate UK. Neither the European Union nor the granting authority can be held responsible for them. UK participation is funded through the UK Horizon guarantee scheme under Innovate UK grant 10093177.
\end{acknowledgements}

\bibliography{Euclid, Q1, mybiblio}

\end{document}